\documentclass[fleqn,usenatbib]{mnras}

\usepackage{graphicx}	
\usepackage{amsmath}	
\usepackage[dvipsnames]{xcolor}  
\usepackage{amssymb}	
\usepackage{newtxtext,newtxmath}    
\usepackage{ulem}
\usepackage{soul}
\usepackage{enumitem}
\usepackage{orcidlink}

\usepackage{macros_vdbosch} 

\begin{document}


\title[Calibrating Core Collapse of SIDM Haloes]
      {Not so Fast: Revised Calibration of SIDM Core Collapse}

\author[van den Bosch \& Tokayer]{%
   Frank~C.~van den Bosch$^1$\orcidlink{0000-0003-3236-2068}\thanks{E-mail: frank.vandenbosch@yale.edu},
Yarone~M.~Tokayer$^2$\orcidlink{0000-0002-0430-5798}
\vspace*{8pt}
\\
   $^1$Department of Astronomy, Yale University, PO. Box 208101, New Haven, CT 06520-8101\\
   $^2$Department of Physics, Yale University, PO. Box 208120, New Haven, CT 06520-8120
}


\date{}

\pagerange{\pageref{firstpage}--\pageref{lastpage}}
\pubyear{2013}

\maketitle

\label{firstpage}


\begin{abstract}
  Self-interacting dark matter (SIDM) haloes undergo gravothermal evolution whose late-time behaviour is central to predictions for dense substructure and black-hole seed formation. We revisit the calibration of this evolution using a new spherical Direct Simulation Monte Carlo code, \SphMCns, which follows spherically symmetric SIDM systems without force softening and to substantially higher central densities than standard $N$-body simulations. We compare \SphMC calculations with conventional $N$-body simulations and with one-dimensional gravothermal fluid models. We show that $N$-body results converge to \SphMC only for softening lengths significantly smaller than those normally adopted for collisionless simulations, and that allowing multiple scatterings per particle per time step does not introduce significant errors when implemented appropriately. Calibrating the gravothermal fluid equations against \SphMCns, we find that the commonly used harmonic interpolation between the long- and short-mean-free-path conductivities is inaccurate. A gentler transition, however, yields excellent agreement for NFW haloes over a range of concentrations and truncation radii, with a single long-mean-free-path conductivity parameter. The revised calibration gives a long-mean-free-path collapse time that is somewhat shorter than previous estimates, but we show that collapse slows dramatically once the core becomes optically thick. We provide fitting functions for the late-time evolution and argue that black-hole seed formation by SIDM gravothermal collapse has generally been underestimated in duration, making it unlikely to explain the massive high-redshift black holes inferred from JWST observations without additional physics.
\end{abstract}




\section{Introduction}
\label{sec:intro}

Self-interacting dark matter (SIDM) remains one of the most widely studied alternatives to collisionless cold dark matter (CDM). In SIDM models, dark-matter particles undergo non-gravitational scattering, allowing heat and momentum transport within haloes while preserving the large-scale phenomenology of $\Lambda$CDM \citep[][]{Spergel.Steinhardt.00, Tulin.Yu.18, Adhikari.etal.25}. Self-scattering initially transports heat from the hotter outer halo into the colder inner region, driving the formation of an approximately isothermal core and reducing the central density. At later times, the direction of net heat flow reverses and the system enters a gravothermal-collapse phase, during which the central region contracts, heats up and becomes denser \citep{LyndenBell.Wood.68, Balberg.etal.02, Koda.Shapiro.11, Outmezguine.etal.23}. This behaviour is closely related to the gravothermal catastrophe long studied in collisional stellar systems. 

Gravothermal collapse has attracted increasing attention because it provides a natural mechanism for generating substantial halo-to-halo diversity. Haloes with similar virial masses but different concentrations, accretion histories or tidal environments need not be observed in the same thermodynamic state: some may still be in the core-expansion phase, while others may already have collapsed to much higher central densities. This offers a plausible route to understanding the observed diversity of rotation curves, including the coexistence of diffuse and comparatively dense systems within the same broad mass range \citep{Ren.etal.19, Sameie.etal.20, Nadler.etal.23, Roberts.etal.25, Fischer.Yu.26}. 

Related considerations arise in the context of strong gravitational lensing. Observational studies have pointed to unexpectedly dense perturbers \citep[][]{Minor.etal.21} and to an excess of galaxy-galaxy strong lensing in clusters relative to standard theoretical expectations \citep{Meneghetti.etal.23, Natarajan.etal.26}. SIDM subhaloes that have undergone gravothermal collapse provide a natural source of compact, high-density structures capable of enhancing such signals, particularly when tidal effects accelerate their evolution after infall \citep{Nishikawa.etal.20, Gilman.etal.21, Zeng.etal.22, Shah.Adhikari.24, Dutra.etal.25}. 

Perhaps the most far-reaching implication of SIDM gravothermal collapse is that sufficiently deep collapse can trigger black hole (BH) formation. In particular, if the self-interactions of the dark matter particles do not allow the formation of any bound states, the runaway gravothermal collapse will drive the core to relativistic velocities at which point the core undergoes catastrophic collapse into a BH via the radial instability \citep[][]{Zeldovich.Podurets.66, Shapiro.Teukolsky.85, Balberg.Shapiro.02}.  More recent studies have revived this idea in several contexts, including intermediate-mass BH formation in Milky Way satellites \citep[][]{Meshveliani.etal.23}, and BH seeding in high-redshift haloes which might help to explain the unexpectedly early appearance of massive accreting BHs in the Universe \citep[][]{Pollack.etal.15, Choquette.etal.19, Feng.etal.21, Feng.etal.22}. This connection has become especially timely in light of the discovery by the James Webb Space Telescope (JWST) of the “little red dots” (LRDs): a numerous population of compact, red, broad-line sources at high redshift that appear consistent with rapidly growing black holes embedded in dense gaseous environments \citep[][]{Kocevski.etal.23, Greene.etal.24, Matthee.etal.24}. Several recent works have therefore proposed SIDM core collapse as a viable channel for producing the heavy black-hole seeds needed for LRDs \citep[][]{Feng.etal.25b, Roberts.etal.25c, Shen.etal.25, Jiang.etal.26}.

In each of these applications, the relevant phenomenology depends sensitively on the gravothermal-collapse time. Whether one is interested in the diversity of dwarf-galaxy rotation curves, the existence of dense strong-lensing perturbers, or the formation of early black-hole seeds, the viability of the SIDM interpretation depends on whether collapse can occur within the available cosmic time. In practice, collapse times are inferred using two main approaches: particle-based $N$-body simulations with stochastic scattering and one-dimensional (1D) gravothermal fluid (GTF) calculations. These methods are complementary, but they rely on different assumptions and each has its own shortcomings.

The gravothermal fluid approach, introduced for SIDM haloes by \citet{Balberg.etal.02} and developed further in many later studies \citep[][]{Ahn.Shapiro.05, Koda.Shapiro.11, Pollack.etal.15, Essig.etal.19, Nishikawa.etal.20, Outmezguine.etal.23}, assumes a spherical, non-rotating halo evolving quasi-statically through a sequence of hydrostatic states. Its main advantage is that it is inexpensive and can follow the system over a large dynamic range, making it well suited for mapping collapse times across parameter space. The key limitation, however, is that one of the main ingredients, the effective conductivity in the long-mean-free path (LMFP) regime, is not know from first principles. Instead, it is typically parameterised by a single free parameter which is calibrated against the results from $N$-body simulations. In addition, one generally interpolates between the long and short-mean-free part (SMFP) regimes using an ad-hoc interpolation scheme that has not been adequately tested.

The alternative route is direct $N$-body simulation, in which SIDM scattering is implemented probabilistically between neighbouring particles on top of the collisionless gravitational evolution \citep{Burkert.00, Kochanek.White.00, Yoshida.etal.00, Dave.etal.01, Vogelsberger.etal.12, Rocha.etal.13}. The major advantage of this approach is that it does not impose spherical symmetry, hydrostatic equilibrium or isotropy, and can naturally capture mergers, cosmological accretion and tidal stripping \citep{Vogelsberger.etal.12, Rocha.etal.13, Zavala.etal.13, Sameie.etal.20, Turner.etal.21, Zeng.etal.22}. This is relevant because the collapse time can be strongly altered by the environment: tidal stripping tends to accelerate collapse by steepening the outer density profile and enhancing the outward heat flux \citep{Nishikawa.etal.20, Sameie.etal.20, Zeng.etal.22, Shah.Adhikari.24}, while continued accretion, mergers or other heating processes can delay or interrupt collapse \citep{Ahn.Shapiro.05, Turner.etal.21, Outmezguine.etal.23, Silverman.etal.26}. On the other hand, properly resolving the gravothermal collapse is numerically challenging. As the core contracts, the central density rises rapidly, local scattering and dynamical time-scales become very short, and the inferred collapse time becomes sensitive to particle number, force softening, time integration, and the details of the scattering implementation \citep{Meskhidze.etal.22, Palubski.etal.24, Fischer.etal.24, Fischer.etal.25, Mace.etal.24}. Recent numerical studies have shown that these effects can shift the collapse time appreciably, so even the simulation-based estimates are not yet fully settled. This in turn impacts the calibration of the LMFP conductivity used in the GTF codes, and thus the collapse times that these codes predict. 

In this work we study the gravothermal core collapse of SIDM haloes using a two-step approach. We first present a new simulation code, similar in spirit to the kinetic solver {\tt KiSS-SIDM} recently developed by \citet{Gurian.May.25} and to the {\tt NSphere} code developed independently by \citet{Kamionkowski.Sigurdson.25a}. This code has several advantages over standard $N$-body codes. Here, we use it  to calibrate the conductivity used in \GTF codes. We show that the harmonic form that has hitherto been used to interpolate the conductivity parameter between the long- and short-mean-free paths is inaccurate. We propose an alternative form and show that a single value for the conductivity parameter in the LMFP regime suffices to model the gravothermal evolution of any NFW halo, independent of its concentration parameter or the level to which it has been truncated. 

This paper is organized as follows. Section~\ref{sec:method} presents the various numerical methods used here to study the gravothermal evolution of SIDM haloes, including a detailed description of the haloes used throughout. Section~\ref{sec:res} starts with a validation of our new \SphMC code against $N$-body simulations, followed by an improved calibration of the conductivity used in the gravothermal fluid equations and a discussion of outstanding issues. In Section~\ref{sec:tccNFW} we present a detailed discussion of the core collapse time of NFW haloes and present a fitting function that accurately captures the late stages of their gravothermal collapse. This is used in Section~\ref{sec:BH} to demonstrate that the time-scale on which core collapse can produce a seed black hole mass is orders of magnitude larger than previously assumed. We summarize our findings in Section~\ref{sec:concl}. Finally, Appendix~\ref{App:Nbodysims} describes the Monte-Carlo method used in our $N$-body simulations, including a detailed demonstration that, contrary to claims in the literature, it is justified to allow multiple scatter events per particle per time. Appendix~\ref{App:realisation} discusses the impact of stochasticity due to realisation noise on the ability to determine the core collapse time of an SIDM halo, and Appendix~\ref{App:validation} presents a validation of the new \SphMC code. Throughout, we assume a flat $\Lambda$CDM cosmology with $\Omega_\rmm=0.3$ and $H_0 =70\kmsmpc$, unless stated otherwise.

\section{Methodology}
\label{sec:method}

The main goal of this paper is to use simulations of SIDM haloes undergoing core collapse to calibrate the conductivity used in gravothermal fluid (hereafter \GTFns) codes to model the same process. For the simulations, we use a new spherical Direct Simulation Monte Carlo code developed by the authors, called \SphMCns, which is similar to, but independent of, two separate codes recently developed by \citet{Gurian.May.25} and by \citet{Kamionkowski.Sigurdson.25a} and~\citet{Kamionkowski.etal.25b}, respectively. 

\SphMC assumes spherical symmetry and integrates the radial coordinates of particles in an $N$-body system following the approach pioneered by H\'enon and Spitzer more than half a century ago \citep[][]{Henon.64, Henon.71, Spitzer.Hart.71}, while treating the collisional scattering using the Direct Simulation Monte Carlo (DSMC) method \citep[][]{Bird.94}. The main advantages of this method are that it (i) is much faster than the $N$-body method, (ii) can integrate the system down to much higher core densities, (iii) does not require any particle softening, and (iv) typically results in much better energy conservation than standard $N$-body codes. The downside is that it only works for spherically symmetric systems. As already advocated by \citet{Gurian.May.25} and \citet{Kamionkowski.etal.25b} this method is ideally suited for studying the core collapse of (spherical) SIDM haloes. 

The \GTF method solves the evolution of a spherical SIDM halo on a 1D radial grid treating the system as a gravothermal fluid following \citet{LyndenBell.Eggleton.80} and \citet{Balberg.etal.02}. The main advantages of this method are that it is even faster than the \SphMC method, and that it is not subject to realisation noise; as input it takes a discretised version of the smooth density and velocity dispersion profiles of the system in question, rather than the phase-space coordinates of a discrete set of points sampled from the system's distribution function. As shown in Appendix~\ref{App:realisation}, suppressing the realisation noise sufficiently for a reliable measure of the core collapse time-scale, $N$-body and \SphMC simulations typically require at least $10^6$ particles per halo \citep[see also][]{Mace.etal.24}.  As a consequence, it typically takes several hours to simulate the core collapse of a single halo with the \SphMC code, and 5-10 times longer with a standard $N$-body code.  The \GTF code, on the other hand, typically already yields converged results for a radial grid consisting of only a few hundred bins and it only takes a few seconds to run it to central densities comparable to what can be achieved with \SphMC or $N$-body simulations. The main downside of the \GTF method is that it has a free parameter that characterises the conductivity in the LMFP regime. This parameter directly impacts the time-scale on which the system undergoes core collapse but has thus far eluded a derivation from first principles. Hence, it is common to calibrate this parameter against $N$-body simulations. However, since the latter are subject to significant uncertainties and systematics \citep[see][and Section~\ref{sec:compA}]{Palubski.etal.24, Fischer.etal.24, Mace.etal.24}, this gives rise to a catch-22 situation. Indeed, different studies have inferred calibration constants that differ by more than 25\% and it is currently unclear whether this reflects inconsistencies among the $N$-body simulations or whether the calibration parameter itself varies with properties of the halo in question. The goal of this paper is to avoid these issues by calibrating the \GTF method against results of the \SphMC method. As we will show, part of the issue is that the ad-hoc interpolation scheme that has been used to link the short- and long-mean-free path regimes is inaccurate.
\begin{figure*}
\centering
\includegraphics[width=\textwidth]{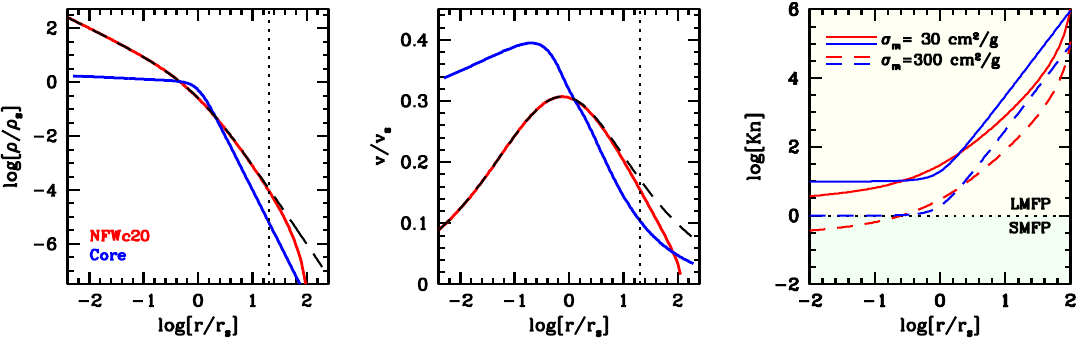}
\caption{Initial density (left) and velocity dispersion (middle) profiles for the truncated NFWc20 halo (red curves) and the \Core halo (blue curves). The black-dashed curves correspond to the untruncated NFW profile and are shown for comparison. Note that the density profile of the truncated NFW halo is almost indistinguishable from that of the untruncated NFW profile inside the virial radius, indicated as the vertical, dotted line. The right panel shows the radial profile of the Knudsen number, $\Kn$, in the initial conditions for two representative values of the collision cross section, as indicated. The horizontal dotted line corresponds to $\Kn=1$, which separates the SMFP regime ($\Kn<1$, indicated in light-green) from the LMFP regime ($\Kn>1$, indicated in light-yellow). See text for discussion.}
\label{fig:initprofs}
\end{figure*}

\subsection{Dark Matter Haloes}
\label{sec:haloes}

We mainly focus on SIDM haloes that start out with an NFW profile
\begin{equation}
\label{NFWprof}
   \rho(r) = {\rho_\rms \over (r/r_\rms) (1 + r/r_\rms)^2} \,,
\end{equation}
\citep{Navarro.etal.97}, with $r_\rms$ and $\rho_\rms$ the characteristic radius and density, respectively. We emphasize that, in the case of SIDM, haloes are only expected to start out with NFW profiles if the collisional relaxation time due to the self-interactions is much longer than the timescale for violent relaxation. This may be violated for some of the more extreme cross sections considered in this study. We emphasize, therefore, that our simulations are to be treated as numerical experiments geared towards improving our understanding of gravothermal collapse. They do not necessarily reflect the most realistic initial conditions.

Throughout, we truncate the initial density profile to assure that the halo has a finite mass and we define the concentration parameter $c \equiv \rvir/r_\rms$. Here, $\rvir$ is the virial radius inside of which the average density is $\bar{\rho} = \Delta_{\rm vir} \, \rho_{\rm crit}$, with $\rho_{\rm crit}$ the critical density for closure. Throughout this paper, we use either $\Delta_{\rm vir}=97$, which is the typical value for a Planck cosmology \citep[][]{Bryan.Norman.98, Planck18}, or $\Delta_{\rm vir}=200$, which is a value that is often adopted when analysing cosmological simulations \footnote{The reason for using two different values for $\Delta_{\rm vir}$ is one of convenience on our part; most of the $N$-body simulations used here adopted the $\Delta_{\rm vir}=97$ definition and were run as part of another project. Rather than rerunning these simulations with $\Delta_{\rm vir}=200$, which is costly, we decided to compare these existing simulations to results from much cheaper \SphMC runs in which we set $\Delta_{\rm vir}=97$. This has no quantitative impact on any of our results.}

We truncate the NFW density profiles using the energy-truncation method of \citet{Drakos.etal.17}\footnote{This is similar to the method used to create King models by modifying the DF of an isothermal sphere \citep[][]{Michie.63, King.66}.} as implemented in the public IC-generation code \PIANISTpy (Chiang et al., in prep.). This method is self-consistent and produces initial conditions that are in perfect equilibrium, contrary to, for example, hard truncation at the virial radius, which results in a dipole instability \citep[][]{Weinberg.23, Dattathri.etal.25a}. Unless stated otherwise, we truncate the haloes such that the total truncated mass is  $M_{\rm tot} = 2.0 \, \Mvir$. Here $\Mvir = M_{\rm NFW}(<\rvir)$ is the mass enclosed by the {\it untruncated} NFW halo inside $\rvir$. As we demonstrate in Section~\ref{sec:densprofdep}, the gravothermal evolution of such a truncated halo is indistinguishable from that of an untruncated halo.

In addition to the NFW haloes described above, we also consider a halo whose density profile is given by 
\begin{equation}\label{abgprof}
    \rho(r) = {\rho_\rms \over (r/r_\rms)^{0.1} \, (1 + [r/r_\rms]^{4})^{0.975}} \,.
\end{equation}
This density profile rapidly transitions from an outer power-law with $\rho \propto r^{-4}$ to an inner core-like region with $\rho \propto r^{-0.1}$. The same halo was used in \citet{Dattathri.etal.25a} and \citet{vdBosch.Dattathri.26} to study core stalling, where it was shown that this system undergoes core collapse much faster than an NFW profile (when expressed in terms of the characteristic collision time). Hence, it is a good example to test whether the results derived here for an NFW profile are applicable to other density profiles. Note that since this profile has a finite mass, there is no need for truncation. In what follows, we refer to haloes with density profiles of equations~(\ref{NFWprof}) and~(\ref{abgprof}) as NFW and \Core haloes, respectively. 

Throughout, we alternate between using physical and dimensionless units. When using physical units, unless stated otherwise, we set the virial mass of the NFW haloes and the total mass of the \Core halo to $10^{12} \msunh$. Using a Hubble parameter $h=H_0/100\kmsmpc = 0.7$, this implies that the fiducially truncated NFW haloes have a total mass of $M_{\rm tot} = 2.86 \times 10^{12} \Msun$, while that of the \Core halo is $1.43 \times 10^{12} \Msun$. Table~1 lists the characteristic densities, radii, velocities, and time-scales (in physical units) for the various haloes used in this study. Note that the scale radius of the \Core halo is tuned such that it has the same characteristic density as the NFWc20 halo. The haloes listed in Table~1 have been used for a wide range of numerical experiments. Throughout the text, we sometimes show results for only one of these halos, often chosen for convenience. At other times, the choice reflects the availability of numerical experiments run as part of previous projects. We have ensured, though, that none of the quantitative results shown are halo-specific.

When using dimensionless units, we set $G = M_\rms = r_\rms = 1$ and express all densities, masses, and radii in units of $\rho_\rms$, $M_\rms$ and $r_\rms$, respectively. Here, $M_\rms \equiv 4 \pi \, \rho_\rms \, r^3_\rms$. For our (truncated) NFW haloes, $M_\rms = \Mvir/f(c)$, where $f(c) = \ln(1+c)-c/(1+c)$, while for \Core haloes $M_\rms = M_{\rm tot}/1.12526$. Velocities are expressed in units of $v_\rms \equiv \sqrt{G M_\rms/r_\rms}$, the collisional cross section per unit mass is expressed in units of $\sigma_\rms = 4 \pi r^2_\rms/M_\rms$, and time is expressed in units of the collisional relaxation time 
\begin{equation}\label{tcoll}
 t_0 = (a \rho_\rms \, \sigma_\rmm \, v_\rms)^{-1} = (a \tilde{\sigma}_\rmm)^{-1} t_{\rm core}\,.
\end{equation}
Here, $\tilde{\sigma}_\rmm = \sigma_\rmm/\sigma_\rms$ is the dimensionless cross section, $t_{\rm core} \equiv r_\rms/v_\rms$ is the core-crossing time, and $a = 4/\sqrt{\pi} \approx 2.26$ for elastic hard-sphere-like interactions as assumed here \citep[][]{Balberg.etal.02}. Hence, a dimensionless cross section of $\tilde{\sigma}_\rmm = 1/a \simeq 0.443$ corresponds to an average of one collision per core crossing.  

Throughout, we exclusively focus on isotropic elastic scattering with a velocity-independent cross section per unit mass given by $\sigma_\rmm$. While more elaborate models for self-interactions are physically attractive \citep[see e.g.,][]{Tulin.etal.13, Robertson.etal.17, Fischer.etal.22, Arido.etal.25}, the main purpose of this paper is to compare core collapse times obtained with different methods, for which a simple isotropic scattering model suffices. Furthermore, \citet{Outmezguine.etal.23} have shown that the temporal evolution of SIDM haloes exhibits an approximate universality that allows velocity-dependent models to be mapped on to velocity-independent models in a well-defined way.

A useful measure to quantify self-interactions is the (local) Knudsen number, $\Kn$, defined as the ratio of the mean free path, $\lambda_{\rm mfp} = (\rho \sigma_\rmm)^{-1}$, and the gravitational scale height (or Jeans length), $H = v / \sqrt{4 \pi G \rho}$, with $v$ the local 1D velocity dispersion. Where $\Kn \ll 1$ the system is in the SMFP regime and can be treated as a standard fluid, while regions with $\Kn \gg 1$ are said to be in the LMFP regime. If the entire system is in the LMFP regime, its gravothermal evolution is self-similar when expressed as a function of $t/t_0$ \citep[][]{Balberg.etal.02, Ahn.Shapiro.05, Koda.Shapiro.11, Yang.etal.23}. 

The red curves in Fig.~\ref{fig:initprofs} show the density and 1D velocity dispersion profiles of the NFWc20 halo, normalised to $\rho_\rms$ and $v_\rms$, respectively.  For comparison, the black-dashed lines show the corresponding untruncated NFW profile, and the vertical dotted line indicates the corresponding virial radius. Note that the truncated profiles look very similar to the untruncated NFW profile out to the virial radius, with $M(<\rvir) \sim 0.953 \Mvir$. The blue curves show the corresponding profiles for the \Core halo. Note that this halo has a steep  gradient in its outer velocity dispersion profile. As shown in \citet{vdBosch.Dattathri.26}, this facilitates a large conductive flux and thus rapid core collapse. The right-hand panel of Fig.~\ref{fig:initprofs} shows the radial profiles of the Knudsen number in the ICs for the NFWc20 and \Core haloes for two different values of the collision cross section, as indicated. Note that halos with both initial profiles start out well in the LMFP regime for $\sigma_\rmm =30 \cmg$, but that their central regions lie in the SMFP regime for $\sigma_\rmm \gta 300 \cmg$.
%
\renewcommand{\arraystretch}{1.05}
\begin{table*}
\label{tab:sims}
\centering
\begin{tabular}{c|cccccccccc}
\hline
Halo & $\Delta_{\rm vir}$ & $c$ & $\Mvir$ & $M_{\rm tot}/\Mvir$ & $M_\rms$ & $r_\rms$ &    $\rho_\rms$     & $v_\rms$ & $t_{\rm core}$ & $\sigma_\rms$ \\
     &          &              &  $[\Msun]$   &                          & $[\Msun]$& $[\kpc]$ &  $[\Msun\pc^{-3}]$ & $[\kms]$ &   $[\Myr]$      &    $[\cmg]$   \\
 (1) & (2) & (3) & (4) & (5) & (6) & (7) & (8) & (9) & (10) & (11) \\     
\hline
\hline
NFWc10       & $200$ &   $10$ & $1.43 \times 10^{12}$ & $2.00$ & $9.60 \times 10^{11}$ & $23.29$ &  $0.0061$ & $420.89$ & $54.10$ & $33.98$ \\
NFWc20       & $200$ &   $20$ & $1.43 \times 10^{12}$ & $2.00$ & $6.83 \times 10^{11}$ & $11.64$ &  $0.0344$ & $502.12$ & $22.68$ & $11.94$ \\
NFWc30       & $200$ &   $30$ & $1.43 \times 10^{12}$ & $2.00$ & $5.79 \times 10^{11}$ &  $7.76$ &  $0.0986$ & $566.41$ & $13.40$ &  $6.25$ \\
NFWc40       & $200$ &   $40$ & $1.43 \times 10^{12}$ & $2.00$ & $5.22 \times 10^{11}$ &  $5.82$ &  $0.2105$ & $620.74$ &  $9.17$ &  $3.91$ \\
NFWc50       & $200$ &   $50$ & $1.43 \times 10^{12}$ & $2.00$ & $4.84 \times 10^{11}$ &  $4.66$ &  $0.3813$ & $668.43$ &  $6.81$ &  $2.69$ \\
Core        & $200$ & $16.3$ & $1.35 \times 10^{12}$ & $1.06$ & $1.27 \times 10^{12}$ & $14.32$ &  $0.0344$ & $617.43$ & $22.68$ &  $9.71$ \\
\hline
NFWc20$^{*}$ &  $97$ &   $20$ & $4.29 \times 10^{9}$  & $1.27$ & $2.05 \times 10^{9}$  &  $2.14$ &  $0.0167$ &  $64.19$ & $32.56$ & $134.10$ \\
NFWc40$^{*}$ &  $97$ &   $40$ & $4.29 \times 10^{9}$  & $1.27$ & $1.57 \times 10^{9}$  &  $1.07$ &  $0.1021$ &  $79.35$ & $13.17$ &  $43.87$ \\
Core$^{*}$  &  $97$ & $16.3$ & $4.05 \times 10^{9}$  & $1.06$ & $3.81 \times 10^{9}$  &  $2.63$ &  $0.0167$ &  $78.93$ & $32.56$ & $109.05$ \\
\hline
\end{tabular}
\caption{Properties of the various haloes used throughout this paper. Column~(1) indicates the name of the halo. Those starting with `NFW' are all truncated NFW haloes, with those starting with `Core' are untruncated and follow the density profile given by equation~(\ref{abgprof}). Columns~(2) and~(3) list the virial overdensity, $\Delta_{\rm vir}$, and the corresponding concentration parameter $c = \rvir/r_\rms$, where $\rvir$ is the virial radius that encloses a mean density equal to $\Delta_{\rm vir}$ times the critical density. Column~(4) lists $\Mvir$, the enclosed mass of the untruncated halo inside $\rvir$. Columns~(5) and~(6) indicate the total mass in units of $\Mvir$ and the characteristic mass $M_\rms$ in Solar units. Columns~(7)-(9) list the characteristic radius $r_\rms$, characteristic density, $\rho_\rms$, and characteristic velocity, $v_\rms$, as defined in the main text. Finally, column~(10) indicates the core-crossing time $t_{\rm core} \equiv r_\rms/v_\rms$, which is characteristic of the dynamical time at the characteristic radius, while column~(11) lists the characteristic cross section, $\sigma_\rms = 4 \pi r^2_\rms/M_\rms$.}
\end{table*}

\subsection{The Spherical Direct Simulation Monte Carlo Code}
\label{sec:SphMC}

The \SphMC code is based on methods devised in the  1960s and early 70s by H\'enon and Spitzer \citep[e.g.,][]{Henon.64, Henon.71, Spitzer.Hart.71} mainly to model the dynamical evolution of spherical star clusters. These methods explicitly assume spherical symmetry throughout the entire evolution of the system, which allows a drastic speed up of the orbit integration compared to standard $N$-body codes. This method was recently revived and adapted to modelling SIDM haloes in \citet{Gurian.May.25} and independently in a recent series of papers by M. Kamionkowski and K. Sigurdson \citep[][]{Kamionkowski.Sigurdson.25a, Kamionkowski.etal.25b, Kamionkowski.Sigurdson.25c}. We have developed a similar code that uses elements from all these previous studies, as described below.

\SphMC takes as input the same initial condition data as standard $N$-body simulations, i.e., the initial Cartesian positions and velocities of the $N_\rmp$ particles that make up the (spherical) system. Each time step, \SphMC first sorts the particles in the order of increasing distance $r_i$ from the centre, where $i=1,2,...,N_\rmp$. Since typically only a few particles change their rank order per time step, we use a standard insertion sort algorithm, which achieves close-to $\calO(N_\rmp)$ performance for nearly-sorted data. Next, it uses a standard kick-drift-kick leapfrog integration scheme to advance the particles a time step $\Delta t$ solving the radial equations of motion given by
\begin{eqnarray}
{\rmd r_i \over \rmd t} & = & v_{r,i} \,,\nonumber \\
{\rmd v_{\rmr,i} \over \rmd t} & = & -{G \, (i-\frac{1}{2}) \, m_\rmp \over r^2_i} + {l_i^2 \over r_i^3}\,.
\end{eqnarray}
Here, $l_i = r_i v_{{\rm t},i}$ is the specific angular momentum of particle $i$ which, in the absence of self-interactions, is an integral of motion, and $v_\rmr$ and $v_\rmt$ indicate the radial and tangential components of the velocity vector. The quantity $1/2$ in the expression for the radial acceleration accounts for the self-gravity of the particles \citep[][]{Spitzer.Hart.71}.

During pericentric passages of particles with small $l$, the equations of motion can become stiff causing the leapfrog integrator to perform poorly. Following \citet{Kamionkowski.Sigurdson.25a} we use Levi-Civita (LC) regularization \citep[][]{LeviCivita.20} to circumvent these problems, albeit using a somewhat different implementation. After each time step, we compute the energy of each particle, $E_i = \frac{1}{2} v_i^2 + \Phi(r_i)$. Here $v_i^2 = v_{\rmr,i}^2 + (l_i/r_i)^2$ and the potential energy is computed using
\begin{equation}
 \Phi(r_i) = -G m_\rmp \left[ \frac{i}{r_i} + \sum_{j=i+1}^{N_\rmp} \frac{1}{r_j}\right]\,.
\end{equation}
If $\delta E_i \equiv \vert [E_i(t+\Delta t) - E_i(t)]/ E_i(t) \vert > \delta_{\rm crit}$ then the time step for particle $i$ is repeated using the LC regularization scheme described in detail in Appendix C.2 of \citet{Kamionkowski.Sigurdson.25a}. If needed, we keep refining the LC time stepping until $E_i(t+\Delta t)$ meets the $\delta_{\rm crit}$-criterion. Throughout, we use $\delta_{\rm crit} = 0.01$.
We have verified that setting $\delta_{\rm crit} = 0.1$ produces results that are indistinguishable. 

Once all particles have successfully been integrated forward in time, particles are scattered using the {\it no-time-counter} (NTC) scheme of DSMC \citep[][]{Bird.94}. Using the list of particles sorted by increasing $r_i$, we collate the first $N_{\rm shell}=4096$ particles into a radial shell and estimate its density using
\begin{equation}\label{rhoshell}
\rho_{\rm shell} = {N_{\rm shell} \, m_\rmp \over \frac{4}{3} \pi (r_{\rm max}^3 - r^3_{\rm min})}\,,
\end{equation}
where $r_{\rm min}$ and $r_{\rm max}$ are the extrema of $r_i$ among the $N_{\rm shell}$ particles. In order to ensure that no heat can be transferred across distances larger than the mean-free-path, we check that $\lambda_{\rm mfp} = 1.0/(\rho_{\rm shell} \sigma_\rmm)$ is larger than the bin width $\Delta r = r_{\rm max} - r_{\rm min}$. If this is not the case, we reduce $N_{\rm shell}$ by a factor of two and try again. We repeat this refinement until $\Delta r < \lambda_{\rm mfp}$, each time recomputing the densities of the refined shell using equation~(\ref{rhoshell}). Following \citet{Gurian.May.25}, we only split shells if they have a minimum of $N_{\rm shell} = 32$ particles\footnote{We have experimented with minimum values for $N_{\rm shell}$ of $16$ and $8$ and find that this has a negligible impact on the final results.}. 

Next, for the particles that are part of the shell we compute the majorant number of trial collision pairs
\begin{eqnarray}
\widehat{N}_{\rm coll} & = & {\frac{1}{2} \, N_{\rm shell} (N_{\rm shell}-1) \, W_\rmp \, \sigma_{\rm coll} \, (v_{\rm rel})_{\rm max} \, \Delta t \over V_{\rm shell}} \nonumber \\
& = & \frac{1}{2} \, (N_{\rm shell}-1) \, \rho \, \sigma_\rmm \, (v_{\rm rel})_{\rm max} \, \Delta t\,.
\end{eqnarray}
Here, $W_\rmp = m_\rmp /m_\chi$ is the number of true (SIDM) particles of mass $m_\chi$ per simulation particle of mass $m_\rmp$, $\sigma_{\rm coll} = m_\chi \sigma_\rmm$ is the collisional cross section, $V_{\rm shell}$ is the shell volume, and $(v_{\rm rel})_{\rm max}$ is an upper bound on the relative velocity between two particles in the shell. In practice, we set $(v_{\rm rel})_{\rm max} = 4 \max(|\vec{v}_i|)$, where the maximum is taken over the simulation particles in the shell.
 
Next, we select a number of potential collision pairs, $N_{\rm coll}$, from $\widehat{N}_{\rm coll}$. We use the integer$+$remainder method \citep[][]{Shoja-sani.etal.26}, which sets $N_{\rm coll} = \lfloor \widehat{N}_{\rm coll} \rfloor$. Here, $\lfloor x \rfloor$ denotes the floor function that returns the largest integer less than or equal to $x$. For each particle, we keep track of a remainder-accumulator, $A_i$, which is initialised to zero at $t=0$ and updated according to 
\begin{equation}
A_i \leftarrow A_i + \frac{(N_{\rm coll} - \lfloor \widehat{N}_{\rm coll} \rfloor)}{N_{\rm shell}}
\end{equation}
If $A_{\rm shell} \equiv \sum_{i=1}^{N_{\rm shell}} A_i > 1$, we set $N_{\rm coll} \leftarrow N_{\rm coll} + 1$ and $A_i \leftarrow A_i - (1/N_{\rm shell})$ for $i=1,..,N_{\rm shell}$. This method ensures that the time-averaged collision rate is exact.

For each of the $N_{\rm coll}$ trial collision pairs, we randomly sample two particles from the shell that we collide if their relative velocity satisfies
\begin{equation}
\calR < {v_{\rm rel} \over (v_{\rm rel})_{\rm max}}
\end{equation}
where $\calR$ is a random deviate from a uniform distribution on $[0,1)$. For each of the colliding particles, $i$ and $j$, we randomly assign an angle $\varphi$, drawn uniformly from $0$ to $2 \pi$, which we use to compute the azimuthal velocity components $v_{\theta} = v_\rmt \cos\varphi$ and $v_{\phi} = v_\rmt \sin\varphi$. We then compute the post-collision velocity components assuming elastic scattering:
\begin{equation}\label{vcoll}
\bv_i = \bv_{\rm cm} + (\bv_{ij}/2) \, \hat{\be} \, \nonumber 
\end{equation}
\vskip -0.55truecm
\begin{equation}
\bv_j = \bv_{\rm cm} - (\bv_{ij}/2) \, \hat{\be} \,,
\end{equation}
where $\bv_{ij} = \bv_j - \bv_i$, $\bv_{\rm cm}$ is the centre-of-mass velocity of the pair, and $\hat{\be}$ is a unit directional vector that we draw randomly from the unit sphere. Finally, we update the angular momenta of the particles using $l \leftarrow r \, v'_\rmt$, where $v'_\rmt$ is the post-collision tangential velocity. Note that this procedure conserves both energy and linear momentum. 

Having completed all pair collisions in the radial shell in question, we select the next
$N_{\rm shell}=4096$ particles from the radius-sorted list, and repeat the above steps. This is repeated until the radial shells have covered all particles. 

All particles are propagated with the same adaptive time step, given by
\begin{equation}
\Delta t = \min(\Delta t_{\rm coll}, \Delta t_{\rm grav})
\end{equation}
where
\begin{equation}\label{dtcoll}
\Delta t_{\rm coll} = \min_j\left({\zeta \over \rho_j \, \sigma_\rmm \, v_{\rm max}}\right)
\end{equation}
is proportional to the minimum collision time among all shells, and 
\begin{equation}\label{dtgrav}
\Delta t_{\rm grav} = \sqrt{2 \, \eta \, \min_i\left({r_i \over |a_r|_i}\right)} 
\end{equation}
is proportional to the minimum dynamical time among all particles, with $a_r$ the radial acceleration. Throughout, we adopt $\zeta = 1$ and $\eta=0.01$, unless explicitly stated otherwise. As discussed below, detailed tests show that this leads to converged results with good energy conservation and scatter rates that are in excellent agreement with expectations. In addition, Appendix~\ref{App:validation} shows that the results obtained with \SphMC are in excellent agreement with those obtained using {\tt KiSS-SIDM}.

\subsubsection{Spatial and temporal resolution}
\label{sec:spatiotemporal}

Similarly to $N$-body simulations, \SphMC simulations are limited by finite spatial and temporal resolution. The finite spatial resolution mainly manifests as an inability to resolve the ever-shrinking mean-free-path of the particles during core collapse. As indicated above, we refine the radial shells in an attempt to ensure that  $\Delta r < \lambda_{\rm mfp}$, but we only  split the shells if they have a minimum of $N_{\rm shell} = 32$ particles. Each halo will, at some point during its core collapse, reach a point when the central region can no longer be refined because of this constraint. We define $r_{\rm fail}$ as the radius within which the refinement criterion can no longer be satisfied. In principle, for $r < r_{\rm fail}$ the system experiences too much transport due to the fact that the simulation fails to resolve the mean-free-path. However, as long as $r_{\rm fail}$ is small compared to the system's core radius this enhanced transport will have little net effect since there are no significant density or temperature gradients within the core. In practice, as detailed in Section~\ref{sec:NFW}, we find that when $r_{\rm fail}$ exceeds the core radius the results start to reveal weak signs of accelerated core collapse compared to the gravothermal fluid simulations due to this artificially enhanced transport. 

The finite temporal resolution mainly comes at play for large collision cross sections. As is evident from equations~(\ref{dtcoll})-(\ref{dtgrav}), the ratio $\Delta t_{\rm coll}/\Delta t_{\rm grav} \propto \Kn$. During core collapse, the Knudsen number of the core continues to decrease. Hence, at some point the time stepping of the simulations is controlled by the collision time rather than the dynamical time, and this transition occurs earlier for larger $\sigma_\rmm$. During core collapse $\Delta t_{\rm coll}$ rapidly decreases with time, and at some point the simulations have to be terminated because progress becomes too slow and computationally expensive.

In principle, the simulations can be greatly sped up using larger values of $\zeta$. We have experimented using values for $\zeta$ as large as $100$, and find that this has a negligible impact on the results; the system's evolution is virtually indistinguishable from that obtained using $\zeta=1$. At first sight this seems surprising, since using a time step $\Delta t \gg t_{\rm coll}$ implies that the particles can `free-stream' too large a distance in between collisions, thus once more causing transport that is too efficient. Note, though, that $t_{\rm coll}$ is defined as the {\it minimum} collision time, which typically occurs at the centre of the core. Outside of the core, both $\rho$ and $v_{\rm max}$ decline rapidly, so that most of the illicit free streaming is confined to the isothermal core, where the net transport is negligible. We suspect that this explains why large values of $\zeta$ do not seem to have significant adverse effects, but we leave a more detailed investigation for future work. In what follows, we take the conservative approach, and adopt $\zeta=1$
throughout.

\subsection{Gravothermal Fluid Code}
\label{sec:gravothermal}

We compare the results of our \SphMC code to those obtained with a gravothermal fluid (\GTFns) code, which treats the SIDM halo as a conducting gravothermal fluid that can be treated as a mono-atomic ideal gas. Originally developed by \citet{LyndenBell.Eggleton.80} to model the collisional evolution of globular clusters, it has since been used by numerous authors to model the evolution of SIDM haloes \citep[e.g.,][]{Balberg.etal.02, Koda.Shapiro.11, Pollack.etal.15, Essig.etal.19, Nishikawa.etal.20}. The \GTF method solves the gravothermal fluid equations on a 1D radial grid, and thus also assumes spherical symmetry.

Consider a spherical SIDM halo with a density profile $\rho(r,t)$ and an enclosed mass $M(r,t)$ at radius $r$ and time $t$. Assume the halo to be isotropic and in quasi-static virial equilibrium with a 1D velocity dispersion $v(r,t)$. Due to self-interactions, characterised by a cross section per unit mass $\sigma_\rmm$, heat flows in the radial direction with a luminosity $L(r,t)$ through a spherical shell at radius $r$. The corresponding heat flux obeys Fourier's law of thermal conduction, 
\begin{equation}\label{thermalconduct}
\frac{L}{4 \pi r^2} = -\kappa \frac{\partial T}{\partial r}\,,
\end{equation}
with $\kappa$ the conductivity and $k_\rmB T = m v^2$  a measure of the temperature of the particles of mass $m$. In addition to the heat conduction equation, the gravothermal fluid also obeys mass conservation,
\begin{equation}\label{massconservation}
\frac{\partial M}{\partial r} = 4 \pi r^2 \rho\,,
\end{equation}
hydrostatic equilibrium, 
\begin{equation}\label{hydrostatic}
\frac{\partial \rho v^2}{\partial r} = -\frac{G M \rho}{r^2}\,,
\end{equation}
where we have defined the pressure $P = \rho v^2$, and the first law of thermodynamics which can be written as
\begin{equation}\label{firstlaw}
\frac{\partial L}{\partial r} = -4 \pi \rho r^2 v^2 \left( \frac{\partial}{\partial t} \right)_M \ln(v^3/\rho)\,,
\end{equation}
with $s = \ln(v^3/\rho)$ an effective entropy.

Strictly speaking, Fourier's law of thermal conduction is valid in the SMFP regime, where $\Kn \ll 1$. In this regime, Chapman-Enskog theory \citep[][]{Lifshitz.Pitaevskii.81} dictates that the conductivity is given by
\begin{equation}\label{condSMFP}
\kappa_{\rm SMFP} = \frac{3}{2} {b \over a} \, n \, k_\rmB {\lambda_{\rm mfp}^2 \over t_\rmr} = \frac{3}{2} \, b \, {k_\rmB\over m} \, {v \over \sigma_\rmm}\,,
\end{equation}
\citep[][]{Balberg.etal.02, Koda.Shapiro.11}, where $n=\rho/m$ is the number density of particles, $b=25\sqrt{\pi}/32$ is the effective impact parameter and $t_\rmr \equiv \lambda_{\rm mfp}/(av)$ is the local collision time with $a = 4/\sqrt{\pi}$ a coefficient relevant for hard-sphere scattering of particles with a Maxwell-Boltzmann distribution \citep[][]{Reif.65}. 
In the LMFP regime no formal expression for the conductivity exists, but \citet{LyndenBell.Eggleton.80} have shown that an empirical relation of the form
\begin{equation}\label{condLMFP}
\kappa_{\rm LMFP} = \frac{3}{2} \beta \, n \, k_\rmB {H^2 \over t_\rmr}
= \frac{3}{2} \,{\beta \over \pi^{3/2}} \, {k_\rmB\over m} \, {\rho v^3 \sigma_\rmm \over G}\,,
\end{equation}
with $\beta$ a free calibration parameter of order unity (see below), can adequately describe the gravothermal collapse of globular clusters. In order to interpolate between these two regimes,
\citet{Balberg.etal.02} adopted the harmonic form
\begin{equation}\label{kappa_fid}
\kappa = \left[ \kappa_{\rm SMFP}^{-1} + \kappa_{\rm LMFP}^{-1}\right]^{-1}
\end{equation}
which has been used by all subsequent SIDM studies based on the gravothermal fluid equations.

Given appropriate boundary conditions, equations~(\ref{thermalconduct})--(\ref{firstlaw}) may be solved from a set of initial radial profiles. We employ a Lagrangian scheme in which each time step consists of two stages: (1) a conduction step in which equations~(\ref{thermalconduct}) and~(\ref{firstlaw}) are solved, and (2) a ``revirialisation'' step, in which hydrostatic equilibrium (equation~[\ref{hydrostatic}]) is restored by adjusting the radii of the Lagrangian mass shells.

During the conduction step, the specific energy $u_i = 3 v^2_i / 2$, of each cell $i$ is updated while the density profile is held fixed. We do so using an implicit scheme, for which
\begin{align}
    {u_i}^{(n+1)} & = {u_i}^{(n)} +\\
 & \Delta t \cdot f\left(u_{i-1}^{(n+1)}, u_i^{(n+1)}, u_{i+1}^{(n+1)}, \rho_{i-1}, \rho_i, \rho_{i+1}\right)\,, \nonumber
\end{align}
where the superscript denotes the time step and ${u_i}^{(n+1)} = {u_i}^{(n)} + \Delta u_i$.  Combining equations~(\ref{thermalconduct}) and~(\ref{firstlaw}) in the quasi-static limit, we linearise for small changes in $u$, and cast the update equations into a tridiagonal form in the variables $\left\{\Delta u_{i-1}, \Delta u_i, \Delta u_{i+1}\right\}$, which we solve using standard linear algebra routines.

The conduction step pushes the system out of hydrostatic equilibrium, after which the system is revirialised by discretizing and linearising equation~(\ref{hydrostatic}), recasting it into tridiagonal form, and solving for the radial displacements $\left\{\Delta r_{i-1}, \Delta r_i, \Delta r_{i+1}\right\}$. This revirialisation step is iterated until $\max_i |\Delta r_i|/r_i < 10^{-12}$, which rarely requires more than three iterations. Finally, the density profile is updated according to equation~(\ref{massconservation}).

In order to ensure that the linear approximations used in the revirialisation step remain accurate, we adjust the time step such that the maximum fractional change in the specific energy remains sufficiently small,
\begin{equation}\label{ducrit}
 \max_i\left( {\vert\Delta u_i \vert \over u_i}\right) < 10^{-4}
\end{equation}

With these choices, our \GTF implementation is both fast and stable, enabling typical systems to be evolved through central density increases in excess of 10 orders of magnitude in only a few minutes on a standard modern processor. We have verified that the resulting evolution accurately matches that obtained using more conservative time step criteria. 

\subsection{Initial Conditions}
\label{sec:ICs}

All haloes simulated in this study are sampled with $N_\rmp = 10^6$ particles, for which we draw phase-space coordinates from the corresponding isotropic distribution function (DF) $f = f(E)$, which is computed using standard Eddington inversion \citep[][]{Eddington.16, Binney.Tremaine.08}, that is,
\begin{equation}
\label{eddington}
    f(E) = \frac{-1}{\sqrt{8}\pi^2} \frac{\rmd}{\rmd E} \int_0^{-E} \frac{\rmd\rho}{\rmd\Phi} \frac{\rmd\Phi}{\sqrt{\Phi-E}}\,.
\end{equation}
Here $\Phi$ is the gravitational potential and $E =\frac{1}{2}v^2 + \Phi$ is the specific energy. All initial conditions (IC) are constructed using the Python code \PIANISTpy (Chiang et al., in prep.), which produces ICs that are perfectly stable if run in isolation in the absence of self-interactions. Appendix~\ref{App:realisation} demonstrates that $N_\rmp=10^6$ is sufficient to suppress realisation noise to an acceptable level \citep[see also][]{Mace.etal.24}.

For the \GTF code, we typically segment each halo into $N=200$ concentric shells, logarithmically spaced between $r_{\rm min} = 0.001 r_\rms$ and $r_{\rm max} = 200 r_\rms$. However, when integrating to central core densities in excess of $\sim 10^{10} \rho_\rms$, we use smaller $r_{\rm min}$ and larger $N$ in order to better resolve the evolving core (see Section~\ref{sec:tccNFW}).  For each shell, we compute the enclosed mass and the radial velocity dispersion by solving the Jeans equation assuming isotropy. These quantities completely specify the initial conditions from which the density and pressure profiles are trivially inferred. 
\begin{figure*}
\centering
\includegraphics[width=0.96\textwidth]{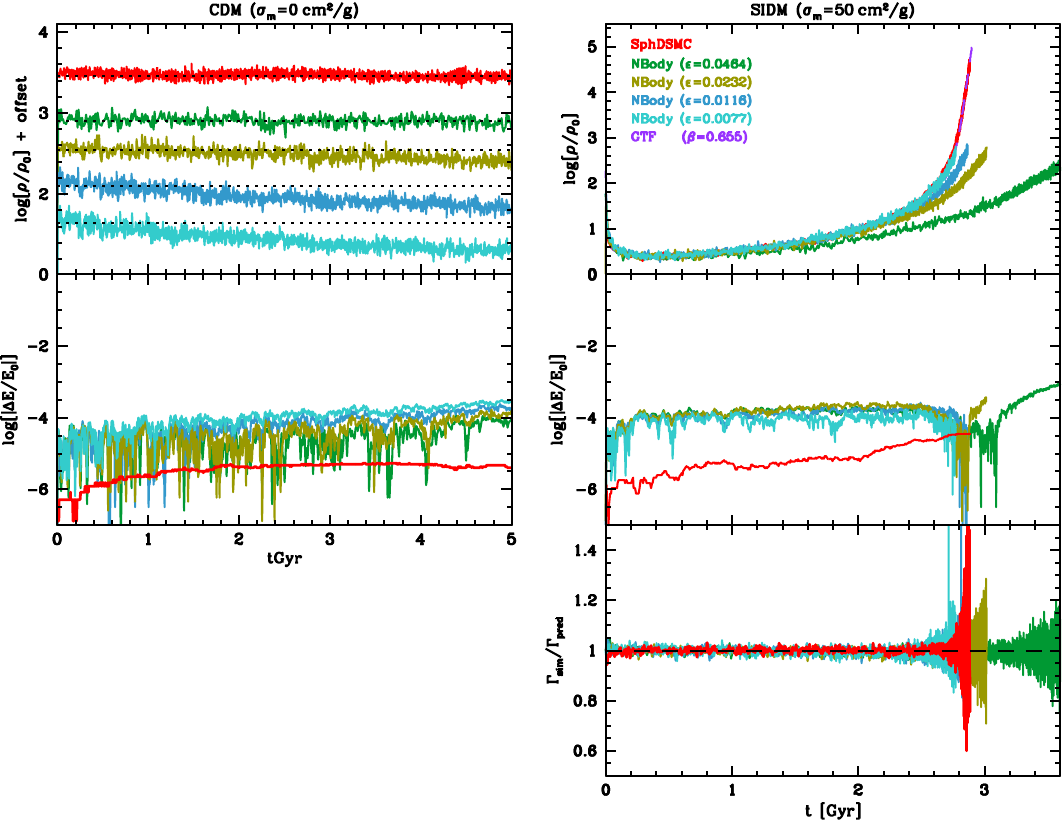}
\caption{Comparison of $N$-body and \SphMC simulations, all with $N_\rmp=10^6$ particles, evolving the NFWc40$^*$ halo assuming CDM (left) and SIDM with $\sigma_\rmm=50\cmg$ (right). Results for the \SphMC simulation are in red, while the green, ocre, azure-blue, and cyan lines show the $N$-body results for four different softening lengths, as indicated. Top panels show the evolution of the central density (as defined in the text). In the top-left panel, results for different simulations have been offset in the vertical direction for clarity, and dotted horizontal lines have been added to guide the eye. In the top-right panel we also overplot the results obtained using the GTF code (in violet) for comparison. The panels in the middle row show the level of energy conservation in the simulations, depicting the temporal evolution of $\vert \Delta E/E_0 \vert = \vert [E(t) - E(0)]/E(0) \vert$ with $E(t)$ the total energy of the halo at time $t$. The lower-left panel shows the ratio of the scatter rate in the simulations to the expected rate computed using equation~(\ref{GammaExp}). See text for a detailed discussion.}
\label{fig:compNFWc40}
\end{figure*}

\subsection{Core Properties}
\label{sec:coreprop}

Throughout, we define the `core' as the region inside the radius $r_\rmc$, defined according to $\rho(r_\rmc) = \rho_0/2$. For the \SphMC simulations, $\rho_0$ is defined as the average density that encloses the innermost 250 particles. For the \GTF code, it is defined as the density of the innermost radial bin. Although these definitions are not identical, in practice the core region is homogeneous enough that we can meaningfully compare these quantities. In addition to the core radius and core density, we also compute the core mass $M_\rmc = M(<r_\rmc)$ and the core's velocity dispersion, $v_\rmc$. The latter is defined as the mass- or particle-weighted velocity dispersion inside $r_\rmc$.

We have also experimented using an alternative definition of the core, for which the core radius is defined as the innermost radius where $\rmd\log\rho/\rmd\log r=-2$. Although this yields values for the core properties that can be quite different during the early stage of core formation, once the system has reached the state in which the core size is largest and the gravothermal collapse kicks in, both core definitions yield comparable results, and none of the results presented here depend significantly on which of these two core definitions we adopt.

For the \SphMC simulations, all radial profiles and their (logarithmic) gradients are computed using Savitsky-Golay filtering \citep[][]{Savitzky.Golay.64} to suppress the impact of discreteness noise.
\begin{figure*}
\centering
\includegraphics[width=\textwidth]{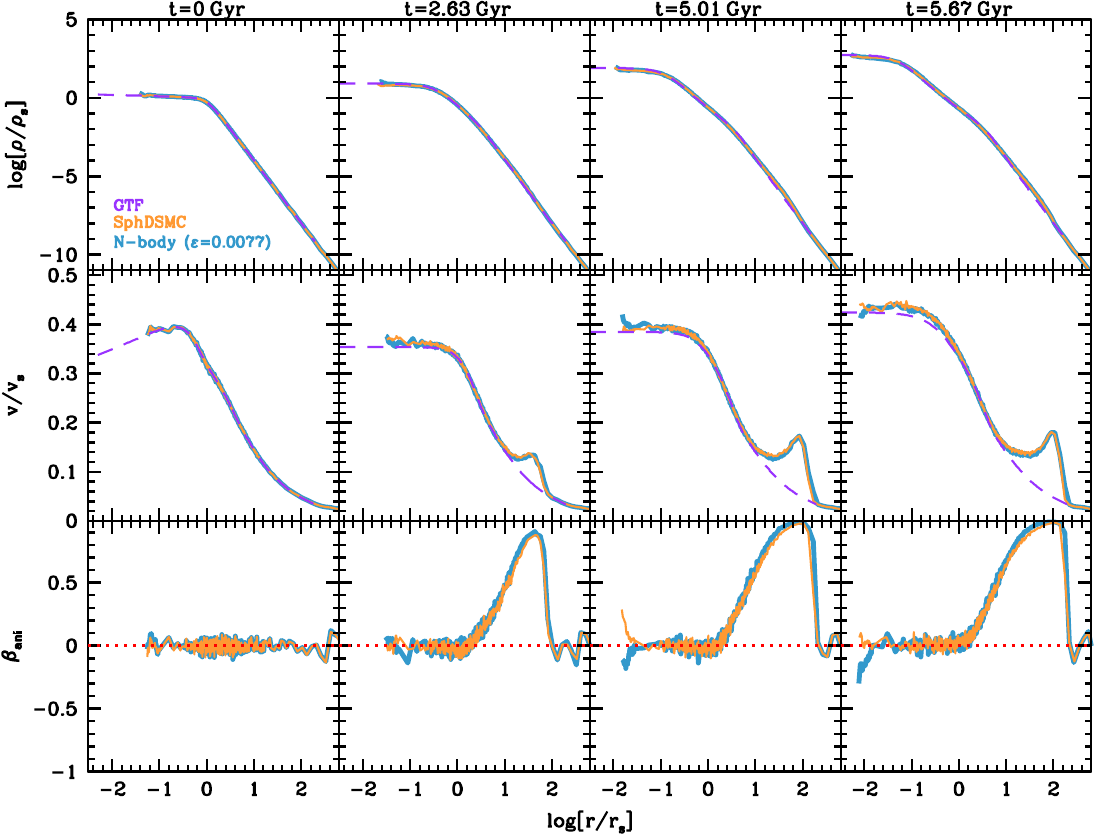}
  \caption{Radial profiles of density, 1D velocity dispersion, and velocity anisotropy of the \Corestar halo (see Table~1) with $\sigma_\rmm = 25 \cmg$ at four different times, as indicated. Each panel compares the results obtained using an $N$-body simulation (in azure-blue), a simulation with the \SphMC code (in orange), and the \GTF code (in violet). Note that the $N$-body and \SphMC results use the same initial condition file with $N_\rmp = 10^6$ particles. Since the \GTF code, which is inherently 1D, cannot capture velocity anisotropy, no \GTF results are shown in the lower panels.}
\label{fig:compABG}
\end{figure*}

\section{Results}
\label{sec:res}

\subsection{Comparing \SphMC to $N$-body}
\label{sec:compA}

Before presenting a detailed comparison of the \SphMC and \GTF codes, we first compare some results of the \SphMC code to those of a standard $N$-body simulation. In particular, we use a modified version of the hierarchical $N$-body {\tt treecode}, written by J. Barnes but modified to account for self-interactions following the method described in detail in \citet[][]{vdBosch.Dattathri.26} to which we refer the reader for more details. Briefly, the code uses a \cite{Barnes.Hut.86} octree to compute accelerations based on a multipole expansion up to quadrupole order and a straightforward second order leap-frog integration scheme to solve the equations of motion. The forces between the particles are softened using a simple Plummer softening, and the opening angle for the tree structure is set to $\theta = 0.7$. The elastic and isotropic scattering of the particles is treated using a Monte Carlo method similar to that of \citet{Vogelsberger.etal.12}. The main difference between our method and that of Vogelsberger et al. and all other methods that have been presented in the literature \citep[e.g.,][]{Kochanek.White.00, Koda.Shapiro.11, Rocha.etal.13, Kim.etal.17, Elbert.etal.18, Fischer.etal.25}, is that we allow particles to undergo multiple scattering events per time step\footnote{While this paper was under review, two papers appeared \citep{Fischer.etal.26, Zier.etal.26} introducing new SIDM implementations in OpenGadget3 and AREPO that also allow for multiple scattering events per time step.}. Appendix~\ref{App:Nbodysims} gives a more detailed description of our scatter treatment including a number of tests that justify this multi-scatter approach.

The top panels of Fig.~\ref{fig:compNFWc40} show the temporal evolution of the central density of the NFWc40$^{*}$ halo\footnote{The asterisks indicates that here we have defined the virial radius as enclosing an average density of $97$, rather than $200$ times the critical density for closure (see Table~1).} in the absence of self-interactions (CDM, top-left) and in the case with $\sigma_\rmm = 50 \cmg$ (SIDM, top-right). The results obtained with the \SphMC code are in red, while the green, ocre, azure-blue, and cyan lines show the results obtained with the $N$-body code for four different values of the (Plummer) gravitational softening length, $\varepsilon$, as indicated. Here and throughout, all softening lengths are expressed in units of the characteristic radius $r_\rms$. 
\begin{figure*}
\centering
\includegraphics[width=0.95\textwidth]{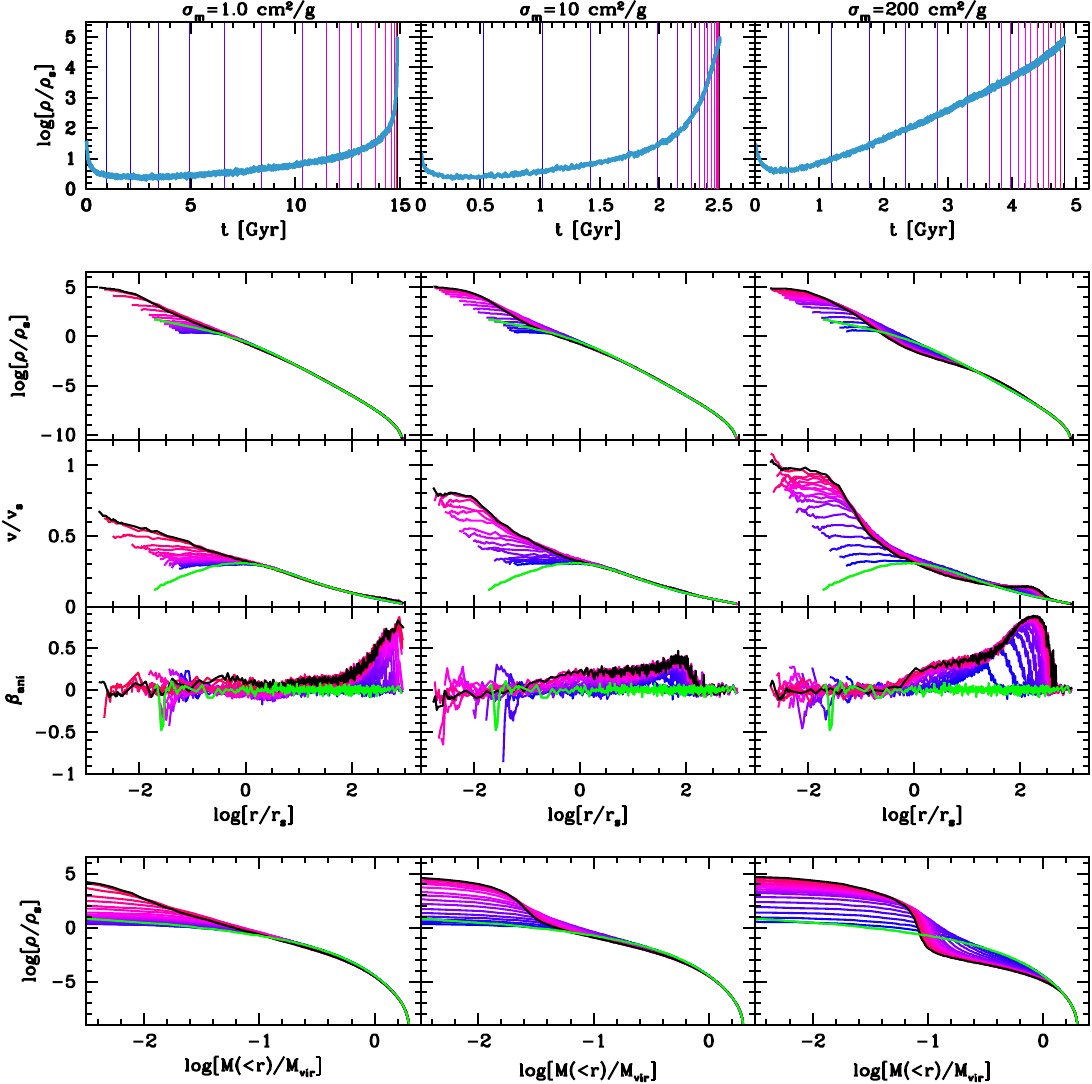}
  \caption{Evolution of the NFWc30 halo for three different values of the collision cross section; $\sigma_\rmm = 1\cmg$ (left), $10\cmg$ (middle) and $200 \cmg$ (right), corresponding to $\tilde{\sigma}_\rmm = 0.16$, $1.6$, and $32$, respectively. Each halo is evolved with the \SphMC code using $N_\rmp=10^6$ particles until the core density reaches $\rho_\rmc = 10^5 \rho_\rms$. The upper panels show the evolution of the core density, nicely depicting the differences as one transitions from core collapse in the LMFP regime (left) to that in the SMFP regime (right). The three middle rows of panels show the radial profiles of density, 1D velocity dispersion, and the velocity anisotropy parameter. The green and black lines show the initial and final profiles, respectively, while  the other coloured lines show the profiles at intermediate times, indicated by vertical lines in the top panels. Finally, the lower panels plot the density at radius $r$ as a function of the enclosed mass within $r$. See text for detailed discussion.}
\label{fig:profsNFW}
\end{figure*}

In the case of CDM, the various lines are offset from each other along the vertical direction for clarity. As is evident, the results for the \SphMC code show that the central density remains constant for the duration of the simulation. This is the expected result, given that CDM is collisionless and that the initial conditions were setup in equilibrium. In the case of the $N$-body simulations, the central density only remains constant in time for $\varepsilon = 0.0464$. For smaller softening lengths, the central density decreases over time, by up to $\sim 60$ percent in $5 \Gyr$ in the case with $\varepsilon = 0.0077$. This is due to insufficient suppression of large-angle scattering between particles, which results in artificial creation of an isothermal core \citep[e.g.,][]{Athanassoula.etal.00, Dehnen.01, Power.etal.03, vdBosch.etal.18b}. 

In the case of SIDM (top-right panel), the \SphMC simulation reveals the characteristic gravothermal evolution of a SIDM halo; initially, the central density rapidly declines as the self-interactions convert the central cusp into an extended isothermal constant-density core due to inward conductive heat transfer. Subsequently, the heat starts to flow outward, from the hot central core to the cooler outskirts, triggering a slow but ever-accelerating increase of the central density characteristic of gravothermal core collapse \citep[][]{LyndenBell.Wood.68, Balberg.Shapiro.02}.

The results for the four $N$-body simulations with different softening lengths clearly diverge at late times, with larger softening length causing core collapse to occur later. This dependence on the choice of softening length has previously been reported by \citet{Mace.etal.24} and~\citet{Palubski.etal.24}. Interestingly, the $N$-body results converge to the results obtained using the \SphMC code, which does not require any softening, in the limit $\varepsilon \to 0.0077$, which is the value for which the CDM simulations reveal strong artificial core formation. We have repeated these tests for (truncated) NFW haloes with different concentrations, and also for the \Core halo, always using $N_\rmp = 10^6$ particles, and consistently find that the $N$-body simulations converge to the results of the \SphMC simulation when $\varepsilon \to 0.0077$, while larger values for $\varepsilon$ result in delayed core collapse. This suggests that SIDM simulations require smaller softening lengths than CDM simulations of the same system to properly resolve core collapse. This notion that SIDM simulations may require a smaller softening length than CDM simulations was also hinted at in \citet[][]{Fischer.etal.25}. It is also consistent with \citet{Mace.etal.24} and~\citet{Palubski.etal.24}, who recommend using a softening length advocated by \citet{vdBosch.etal.18b} for resolving the tidal evolution of (CDM) subhalos, which is significantly smaller than the softening length that is considered optimal for evolving CDM haloes in isolation \citep[for example, as advocated in][]{Power.etal.03}.

It is worth pointing out that the $N$-body simulations are terminated when the central density reaches $10^3\rho_\rms$. This is done because the simulations, all of which are run on a single core without hierarchical time stepping, become unacceptably slow once the core density exceeds this value. However, with the \SphMC simulations, which are significantly faster, we can afford to run the simulations to much higher density. For comparison, while the $N$-body simulation with $\varepsilon=0.0077$ took $175$ hours of wall clock time to complete, the \SphMC simulation only required $4.7$ hours despite continuing to a central density that is a factor 100 higher. Finally, we emphasize that we are not advocating $\varepsilon=0.0077$ as the optimal softening when simulating SIDM in general; we have only demonstrated that this yields the best agreement with the \SphMC results when simulating down to a central density of $\sim 1000 \rho_\rms$. It is to be expected that if we were able to run the $N$-body simulations to even higher densities even smaller softening lengths would be required to resolve the central core. Hence, when using a simulation code that requires force softening to study gravothermal collapse deep into the collapse phase, adaptive softening schemes are likely to be required. This highlights an important advantage of the \SphMC code, which is free of the burden to soften gravity.

The panels in the middle row of Fig.~\ref{fig:compNFWc40} show the  evolution of $\vert \Delta E/E_0 \vert = \vert [E(t) - E(0)]/E(0) \vert$ where $E(t)$ is the total energy of the system at time $t$. Note that energy conservation in the $N$-body simulations is typically at the level of $10^{-4}$, with little to no dependence on $\varepsilon$, and is only slightly worse in the case of SIDM than in the case of CDM. As has been pointed out in several recent studies \citep[][]{Palubski.etal.24, Mace.etal.24, Fischer.etal.24}, achieving such a level of energy conservation with SIDM simulations is extremely challenging when using hierarchical time stepping and/or multiple processors and typically requires extremely small time steps. We are able to avoid these issues by having all particles use the same (adaptive) time step and running the simulations on a single core \citep[see][for a detailed discussion]{Robertson.17}. Note that energy conservation in the \SphMC simulations is more than an order of magnitude better than in the $N$-body simulations. Given that the \SphMC simulations run in a fraction of the time, do not require any force softening, and are able to follow the evolution to significantly higher central densities, it is clear that the \SphMC code is vastly superior to the standard $N$-body approach for studying the gravothermal evolution of SIDM haloes, as long as one is willing to accept the restriction of spherical symmetry \citep[see also][]{Gurian.May.25, Kamionkowski.etal.25b} 

Finally, the lower-right panel of Fig.~\ref{fig:compNFWc40} shows the ratio of the scatter rate in the simulation to the expected rate. The latter is computed using
\begin{equation}\label{GammaExp}
\begin{split}
\Gamma_{\rm pred} &= \frac{1}{2} \int n(\bx) \, \rho(\bx) \, \sigma_\rmm \, \langle v_{\rm rel} \rangle(\bx) \, \rmd^3\bx \\
&= 8 \sqrt{\pi} \, N_\rmp \, \sigma_\rmm \, \frac{1}{M_{\rm tot}} \int_0^{\infty} \rho^2(r) \, v(r) \, r^2 \, \rmd r\,,
\end{split}
\end{equation}
where $n(\bx) = \rho(\bx)/m_\rmp$ is the number density of particles, $v(\bx)$ is the local 1D velocity dispersion, and in the second step we have assumed spherical symmetry and that the particles follow a Maxwell-Boltzmann distribution. We evaluate $\Gamma_{\rm pred}$ every few time steps using the instantaneous density and velocity dispersion profiles of the halo computed on the fly in the simulation codes, and compare its value against the actual scatter rate in the simulation. The latter is evaluated as $\Gamma_{\rm sim} = (\sum_i N_{{\rm scat},i})/\Delta t$, where $N_{{\rm scat},i}$ is the number of scatter events that particle $i$ experienced during the time period $\Delta t$. Note that the scatter rates in both the $N$-body and \SphMC codes are in excellent agreement with the theoretical predictions to within the shot noise\footnote{This noise becomes large during the later stages of the core collapse as a consequence of the fact that the time steps become progressively smaller.}, justifying the Monte-Carlo scatter algorithms in both codes. 


As an additional validation of the \SphMC code, we simulate the gravothermal evolution of the \Corestar halo with $\sigma_\rmm = 25 \cmg$ using both the $N$-body code and the \SphMC code. Fig.~\ref{fig:compABG} shows snapshots of the density profiles (top panels), and the radial profiles of the 1D velocity dispersion 
\begin{equation}
v = \sqrt{(\sigma^2_\rmr + \sigma^2_\rmt)/3}
\end{equation}
with $\sigma^2_x = \langle v_x^2 \rangle - \langle v_x \rangle^2$ (middle panels) and the velocity anisotropy parameter 
\begin{equation}
\beta_{\rm ani} \equiv 1 - {\sigma^2_\rmt \over 2 \sigma^2_\rmr}\,, 
\end{equation}
(bottom panels). Results are shown at four different times, as indicated. Note the excellent agreement between the $N$-body results (blue curves) and those obtained using the completely independent \SphMC code (orange curves). Also note the development of a large `bump' in the velocity dispersion profile at large radii. This feature is due to particles that receive large velocity kicks due to scattering in the dense central region that propels them to highly radial orbits \citep[see also][]{Giersz.Heggie.94, Takahashi.95, Koda.Shapiro.11, Fischer.etal.25, Gurian.May.25}. This is supported by the $\beta_{\rm ani}$ profiles, which show that the halo outskirts become highly radially anisotropic. The core itself, on the other hand, remains, at all times, perfectly isotropic ($\beta_{\rm ani} = 0$), as expected for elastic short-range interactions. For comparison, the violet dashed curves show the predictions obtained using the \GTF code with the best-fit conductivity model presented in Section~\ref{sec:densprofdep}. Although the \GTF results accurately match the density profiles at all times, they do not reproduce the bump-feature in the velocity distribution. This is a consequence of the fact that the \GTF code is inherently 1D and therefore implicitly assumes velocity isotropy throughout. 

\subsection{Gravothermal evolution in the long and short mean free path regimes}
\label{sec:profs}

We now use the \SphMC code to study the gravothermal evolution of SIDM haloes in both the long and the short mean free path regimes.  Fig.~\ref{fig:profsNFW} shows the evolution of a NFWc30 halo for three different values of the collision cross section; $\sigma_\rmm = 1\cmg$, $10\cmg$ and $200 \cmg$, as indicated. This corresponds to dimensionless cross sections of $\tilde{\sigma}_\rmm = 0.16$, $1.6$, and $32$, respectively. Each halo is evolved with the \SphMC code using  $N_\rmp=10^6$ particles until the core density reaches $\rho_\rmc = 10^5 \rho_\rms$. The upper panels show the evolution of the core density. For $\sigma_\rmm = 1\cmg$ (left panels), the evolution is characteristic of gravothermal evolution in the LMFP regime; rapid core formation, followed by a slow, but ever accelerating core collapse. During the final stages, the increase in the core density is catastrophic (i.e., faster than exponential). For $\sigma_\rmm = 10\cmg$ (middle panels), the core collapse is more gradual; although the time to reach $\rho_\rmc = 10^5 \rho_\rms$ is roughly six times shorter than in the case with $\sigma_\rmm=1\cmg$, the rate at which the core density is increasing at that stage is significantly slower. This trend continues to larger cross sections; for $\sigma_\rmm = 200 \cmg$ (right panels), the entire region inside $3.5 r_\rms$ is already in the SMFP regime (i.e., $\Kn < 1$) from the start of the simulation. Following core formation, $\rho_\rmc$ grows (only) at a close-to-exponential rate, and it actually takes almost twice as long to reach a core density of $10^5 \rho_\rms$ as in the case with the 20-fold smaller cross section.
\begin{figure}
\centering
\includegraphics[width=0.45\textwidth]{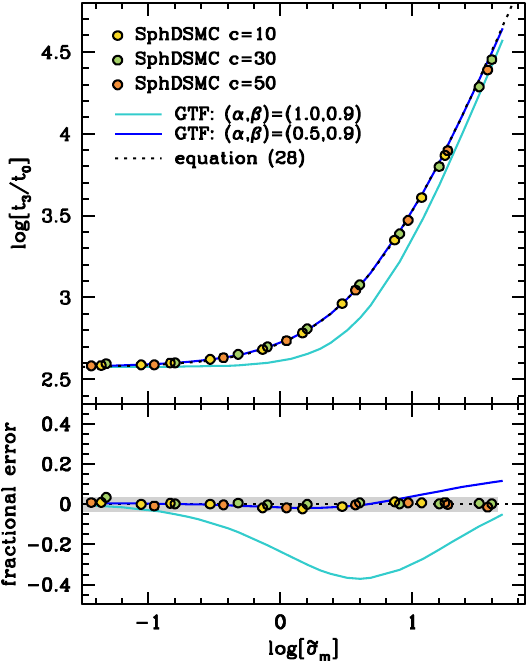}
  \caption{The upper panel plots the ratio $t_3/t_0$ as a function of the dimensionless cross section, $\tilde{\sigma}_\rmm$. Coloured circles correspond to the results from the \SphMC simulations for different values of the halo concentration parameter, as indicated. Blue and cyan lines are predictions obtained with the \GTF code for different combinations of the parameters $\alpha$ and $\beta$, as indicated. The black dotted line shows the results for the fitting function given by equations~(\ref{tnfitNFW}) with the best-fit parameters of Table~2. The lower panels shows the residuals with respect to this fitting function, with the grey-shaded band indicating the  $[-2\sigma,+2\sigma]$ band, where $\sigma$ is the standard deviation in collapse times due to realisation noise (see Appendix~\ref{App:realisation}).}
\label{fig:tcct0}
\end{figure}

The notion that core-collapse behaves differently in the long- and short mean free path regimes has previously been emphasized and demonstrated using gravothermal fluid codes \citep[][]{Balberg.Shapiro.02, Balberg.etal.02, GadNasr.etal.24}. However, to our knowledge, it has never been demonstrated or confirmed in numerical simulations, with the exception of \citet{Koda.Shapiro.11} and \citet{Mace.etal.25} who ran simulations with $\tilde{\sigma}_\rmm$ values as large as unity and noted that such systems already deviate from the self-similar evolution characteristic of the LMFP regime (see discussion at the end of Section~\ref{sec:NFW}). The main reason why most studies based on $N$-body simulations have focussed on small values of $\tilde{\sigma}_\rmm$ is that it is extremely expensive to evolve a system in the SMFP regime. While this remains true for the \SphMC code, it is still feasible to evolve systems with moderate $\tilde{\sigma}_\rmm$ (up to $\sim 30$) up to relatively high densities. 

The reason for the different behaviour in the long and short mean free path regimes is well understood \citep[][]{Balberg.etal.02}. Once the core becomes optically thick, i.e., when the mean free path $\lambda_{\rm mfp}=1/(\rho \sigma_\rmm)$ becomes smaller than the size of the core, it can only transfer heat to the envelope at its surface boundary. Prior to this, the entire core can directly transfer heat, resulting in much more efficient heat transport. This is also evident from the expressions for the conductivities; $\kappa_{\rm LMFP} \propto \sigma_\rmm$ (equation~[\ref{condLMFP}]) while $\kappa_{\rm SMFP} \propto 1/\sigma_\rmm$ (equation~[\ref{condSMFP}]). 

The three middle rows of panels of Fig.~\ref{fig:profsNFW} show the radial profiles of density, 1D velocity dispersion, and the velocity anisotropy parameter, respectively. The green and black lines show the initial and final profiles. The other coloured lines show the profiles at intermediate times, indicated by vertical lines in the top panels. Note how the initial cusp is first converted to an isothermal core, after which the core contracts, increasing its density and central velocity dispersion. Although, by construction, the final density profiles all have $\rho_\rmc = 10^5 \rho_\rms$, the detailed profiles show a clear dependence on $\sigma_\rmm$. In particular, when $\sigma_\rmm$ is large, the density at intermediate radii (roughly around $r/r_\rms=1$) ends up being significantly lower than its initial value. The velocity dispersion profiles also show clear differences, with the core ending up hotter for larger cross sections, even though the core densities at the end are identical. Similarly to the case of the \Corestar halo shown in Fig.~\ref{fig:compABG}, over time a `bump' appears in the velocity dispersion profile at larger radii, albeit much weaker this time; it is most clearly evident in the case of $\sigma_\rmm = 200\cmg$. 

Finally, the lower panels of Fig.~\ref{fig:profsNFW} plot the density at radius $r$ as a function of the enclosed mass within $r$. This nicely illustrates the different core structures in the LMFP and SMFP regimes. In particular, in the latter, one can infer a clear core-within-a-core structure \citep[see also][]{Balberg.etal.02, Feng.etal.21}. This inner core, which becomes denser over time, is entirely in the SMFP regime and only transfers heat to the outer core structure from its surface. The vast majority of heat flux is highly localised at this boundary of the inner core. As shown in \citet{Balberg.etal.02}, in this regime the core mass therefore decreases much slower than in the LMFP regime \citep[see also][]{GadNasr.etal.24}. 

\subsection{The Gravothermal evolution of NFW haloes}
\label{sec:NFW}

We now compare the results obtained with our \SphMC code with those obtained using the \GTF code. As mentioned in Section~\ref{sec:intro} this requires a calibration of the conductivity in the LMFP regime, which is characterised by the free parameter $\beta$ (cf. equation~[\ref{condLMFP}]).

It is common to calibrate $\beta$ by comparing the time-scales for collapse. Unfortunately, when $\sigma_\rmm$ is large and core collapse occurs largely in the SMFP regime, there is no well-defined time of core collapse, given that $\rmd\log\rho_\rmc/\rmd\log t$ is much smaller (cf. upper-right panel of Fig.~\ref{fig:profsNFW}). Therefore, we instead calibrate the \GTF model using the time $t_3$, defined as the time when the core density\footnote{We have also performed this analysis using the alternative definition of core density, $\rho_{-2}$ (see Section~\ref{sec:coreprop}), which yields similar results.} $\rho_\rmc$ reaches a value of $10^3 \rho_\rms$. In what follows, we will loosely refer to $t_3$ as the collapse time.

We use the \SphMC code to evolve a set of spherical isotropic NFW haloes with different concentrations (haloes NFWc10, NFWc30 and NFWc50 listed in Table~1) and different collision cross sections. Each halo has $10^6$ particles and is truncated such that $M_{\rm tot} = 2 \Mvir$. The coloured dots in the upper panel of Fig.~\ref{fig:tcct0} show the ratio $t_3/t_0$, where $t_0$ is the collisional relaxation time given by equation~(\ref{tcoll}), as a function of the dimensionless collision cross section $\tilde{\sigma}_\rmm \equiv \sigma_\rmm/\sigma_\rms$. Different colours correspond to different halo concentrations, as indicated. Note that $t_3/t_0$ is roughly constant for $\tilde{\sigma}_\rmm \lta 0.1$, but increases rapidly for larger cross sections. Importantly, the trend is independent of halo concentration. Next, we run the \GTF code for the NFWc30 halo (results for other halo concentrations are identical) on a fine grid of $\tilde{\sigma}_\rmm$. We initially adopt the standard interpolation scheme for conductivity, given by equation~(\ref{kappa_fid}), and tune $\beta$ to fit the results for small $\tilde{\sigma}_\rmm$ (i.e., in the LMFP regime). This results in a best-fit value of $\beta = 0.90$, which produces the predictions for $t_3/t_0$ shown as the cyan line in Fig.~\ref{fig:tcct0}. As is evident, this drastically underestimates $t_3/t_0$ for $\tilde{\sigma}_\rmm \gta 0.1$. 
\begin{figure}
\centering
\includegraphics[width=0.45\textwidth]{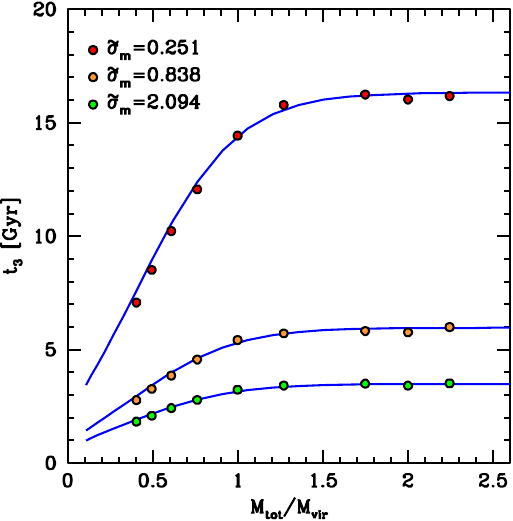}
  \caption{The core collapse time $t_3$ as a function of the total mass of the truncated NFWc20 halo, $M_{\rm tot}$, in units of $\Mvir$. Smaller values of $M_{\rm tot}/\Mvir$ imply a smaller truncation radius, while an untruncated NFW halo has $M_{\rm tot}/\Mvir \to \infty$. Coloured dots indicate results obtained using \SphMC simulations with $N_\rmp=10^6$ for three different values of the collision cross section, as indicated. Blue
  lines are predictions obtained using the \GTF code with $(\alpha,\beta)=(0.5,0.0.90)$, which accurately match the simulation results for any level of truncation.}
\label{fig:tcTrunc}
\end{figure}

In principle, for each $\sigma_\rmm$ one could find a different value for $\beta$ for which the GTF code reproduces the results of the \SphMC simulation, and this is largely what has been done in the literature. We find that the range of $\beta$ values needed to fit the \SphMC simulation results\footnote{For $\tilde{\sigma}_\rmm \lesssim 2$, $t_3 \propto \beta^{-1}$, but for larger cross sections the dependence is weaker, roughly becoming $t_3 \propto \beta^{-0.1}$ for $\tilde{\sigma}_\rmm = 40$} ranges from $0.90$ for $\tilde{\sigma}_\rmm \lta 0.1$ to $\sim 0.4$ for $\tilde{\sigma}_\rmm = 4$. Interestingly, this covers pretty much the entire range of values for $\beta$ that have been obtained by fitting results obtained with GTF codes to $N$-body simulations of NFW haloes, which include values for $\beta$ ranging from $0.9$ \citep[][]{Zhong.etal.23, Mace.etal.25} to $0.6$ \citep[][]{Essig.etal.19, Outmezguine.etal.23}. The lack of even smaller values of $\beta$ is likely the result of the fact that they correspond to relatively large values of $\tilde{\sigma}_\rmm$, which have not been considered for $N$-body simulations given that the CPU requirements are too demanding.

Comparing equations~(\ref{condSMFP}) and~(\ref{condLMFP}), it is clear that $\beta$ is some measure of the effective impact parameter. Hence, while it is expected that $\beta$ depends on the detailed density and/or velocity profile of the halo in question (see Section~\ref{sec:densprofdep}), it makes little sense that it would depend on the collision cross section. Hence, the discrepancy for $\tilde{\sigma}_\rmm \gta 0.1$ is unlikely to be related to $\beta$, but rather to the functional form used to interpolate the conductivity parameter between the LMFP and SMFP regimes.  As discussed in Section~\ref{sec:gravothermal}, virtually all studies of SIDM haloes based on the gravothermal fluid equations have adopted the interpolation scheme of equation~(\ref{kappa_fid}) originally introduced by \citet{Balberg.etal.02}. Here, we generalize this interpolation scheme according to
\begin{equation}\label{kappa_int}
\kappa = \left[ \kappa_{\rm SMFP}^{-\alpha} + \kappa_{\rm LMFP}^{-\alpha}\right]^{-1/\alpha}
\end{equation}
where we treat $\alpha$ as an additional free parameter. We have experimented with a range of $\alpha$ values, each time keeping $\beta$ fixed to $0.90$. The blue curve in Fig.~\ref{fig:tcct0} shows the results for $\alpha = 0.5$, which is an excellent match to the data. Hence, using a simple modification of the interpolation scheme, the \GTF code accurately reproduces the simulation results for all NFW haloes, regardless of $\sigma_\rmm$ or halo concentration. 

For completeness, we point out that early work by \citet{Koda.Shapiro.11} already noted a discrepancy in the gravothermal evolution of SIDM haloes in the SMFP regime between $N$-body simulations and predictions based on the gravothermal fluid equations. However, rather than modifying the interpolation scheme, they suggested modifying the value of the effective impact parameter $b$ that appears in the expression for the conductivity in the SMFP regime given by equation~(\ref{condSMFP}). In particular, they suggest setting $b=0.25$ in the transitional regime where $\Kn \sim 1$, as opposed to adopting $b=25\sqrt{\pi}/32 \approx 1.385$ as appropriate for the SMFP regime \citep[][]{Chapman.Cowling.70, Lifshitz.Pitaevskii.81}. More recently, \citet{Mace.etal.25} noticed that reproducing their SIDM core collapse $N$-body simulations with predictions from a \GTF code requires making the parameter $\beta$ dependent on the cross section, qualitatively in agreement with the results shown in Fig.~\ref{fig:tcct0}. They advocate the use of an `effective' conductivity parameter $\beta_{\rm eff}(\tilde{\sigma}_m)$, defined such that the conductivity $\kappa = \kappa_{\rm LMFP}$, with $\kappa_{\rm LMFP}$ given by equation~(\ref{condLMFP}) but with $\beta$ replaced by $\beta_{\rm eff}$. Hence, they attempt to characterise the conductivity in the SMFP using the expression for the LMFP regime by making the parameter $\beta$ dependent on the cross section. They even advocate using a more general interpolation scheme of the form of equation~(\ref{kappa_int}), which they use to derive the functional form of $\beta_{\rm eff}(\tilde{\sigma}_m)$ and then tune $\alpha$ and $\beta$ to match their $N$-body results. While these modifications suggested by \citet{Koda.Shapiro.11} and \citet{Mace.etal.25} achieve the same goal (tuning \GTF results to match simulation results), we consider our solution of modifying the interpolation scheme more elegant. In addition, neither \citet{Koda.Shapiro.11} nor \citet{Mace.etal.25} included simulations with $\tilde{\sigma}_\rmm > 1$, which limits their ability to probe the SMFP regime. In contrast, our analysis extends all the way to $\tilde{\sigma}_\rmm \approx 20$, allowing for a calibration across a much larger dynamic range.

\subsection{Dependence on density profile}
\label{sec:densprofdep}

In the previous section, we have shown that the \GTF code accurately describes the gravothermal evolution of an SIDM halo with an initial NFW density profile, for any collision cross section, as long as the conductivity is modelled using $(\alpha,\beta) = (0.5, 0.9)$. We now examine whether the same conductivity model also works for haloes with other initial density profiles.

We first consider NFW profiles that are truncated differently. Until now, we have exclusively focused on haloes that were truncated such that $M_{\rm tot}/\Mvir = 2$. The coloured dots in Fig.~\ref{fig:tcTrunc} show the collapse time, $t_3$, obtained with \SphMC simulations, for the NFWc20 halo as a function of the ratio $M_{\rm tot}/\Mvir$. Smaller values of $M_{\rm tot}/\Mvir$ imply a truncation at smaller radii, whereas $M_{\rm tot}/\Mvir \to \infty$ for an untruncated NFW profile. Note that all haloes have $N_\rmp = 10^6$ particles, independent of their level of truncation. Different colours correspond to different values of the collision cross section, as indicated. Note that more truncated haloes undergo core collapse sooner, in agreement with \citet{Nishikawa.etal.20}, and that the collapse time asymptotes to a constant value for $M_{\rm tot}/\Mvir \gta 1.5$. Hence, the results for $M_{\rm tot}/\Mvir = 2$ shown in the previous sections are also applicable to untruncated NFW haloes.
\begin{figure}
\centering
\includegraphics[width=0.45\textwidth]{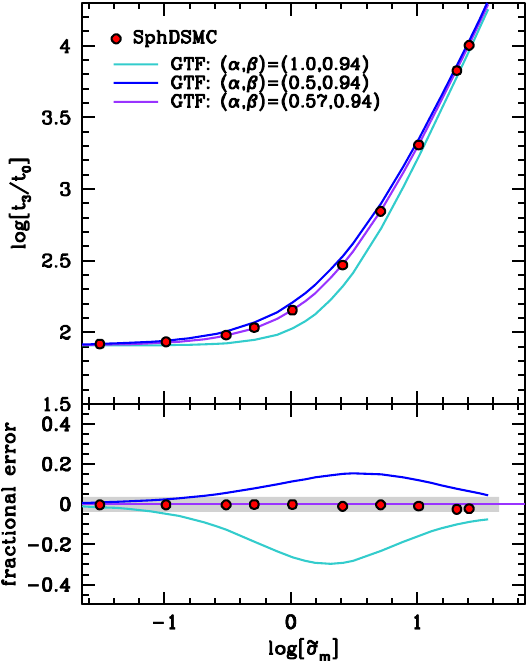}
  \caption{Same as Fig.~\ref{fig:tcct0}, but for the \Core halo. Accurately fitting the \SphMC simulation data (red dots) requires a small adjustment to the parameters $\alpha$ and $\beta$ that quantify the conductivity in the \GTF model, compared to the parameters that best-fit the results for NFW haloes. The residuals in the lower panel are with respect to the \GTF predictions.}
\label{fig:t3ABG}
\end{figure}

The blue curves indicate the predicted collapse times obtained by using the \GTF code with $(\alpha,\beta) = (0.5, 0.9)$. As is evident, this adequately matches the simulation results. Although the fit is not perfect, with the \GTF model slightly overpredicting $t_3$ for strongly truncated haloes and small $\tilde{\sigma}_\rmm$, the discrepancy is only at the few percent-level. Hence, we conclude that the fiducial conductivity model with  $(\alpha,\beta) = (0.5, 0.9)$ accurately accounts for the gravothermal collapse of {\it any} NFW halo, independent of halo concentration or the level by which the halo has been truncated. 
\begin{figure*}
\centering
\includegraphics[width=0.95\textwidth]{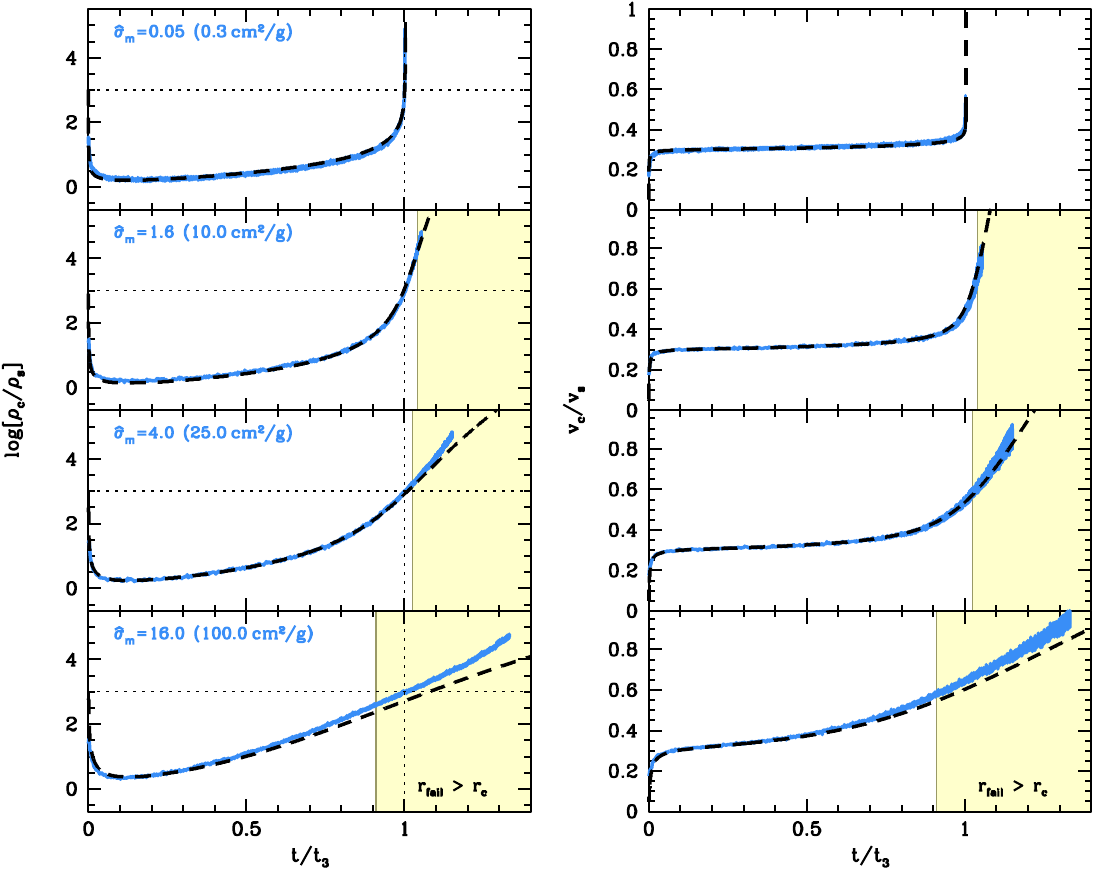}
  \caption{The evolution of the core density (left panels) and core velocity dispersion (right panels) of the NFWc30 halo as a function of time, normalised by $t_3$, for four different values of the (dimensionless) cross section $\tilde{\sigma}_\rmm$, as indicated. The azure-blue curves show the results obtained using \SphMC simulations with $N_\rmp=10^6$ particles, while the black-dashed lines are the predictions from the \GTF code using the new calibration of the conductivity with $(\alpha,\beta)=(0.5,0.9)$.  Overall, the \GTF predictions are in good agreement with the \SphMC results, except in the yellow shaded regions, which indicate the times when $r_{\rm fail}>r_{\rm core}$ in the \SphMC simulations (see text for details). Some curves appear to become thicker at later times; this reflects fluctuations in the values of $\rho_\rmc$ and $v_\rmc$, mainly due to discreteness noise, at subsequent outputs.}
\label{fig:evolNFW}
\end{figure*}

Next, we move away from NFW haloes and focus instead on the \Core halo, whose density profile is given by equation~(\ref{abgprof}). The key difference with respect to an NFW halo is that the \Core profile has a central $r^{-0.1}$ core, as opposed to a $r^{-1}$ cusp. In addition, the outer density profile falls of as $r^{-4}$, as opposed to $r^{-3}$ in the case of an untruncated NFW profile. This implies that the system has finite mass and that no truncation is required.  The density of the \Core halo is normalised such that it has the same characteristic density $\rho_\rms$ as the fiducial NFWc20 halo (see Table~1). 
\begin{figure*}
\centering
\includegraphics[width=0.9\textwidth]{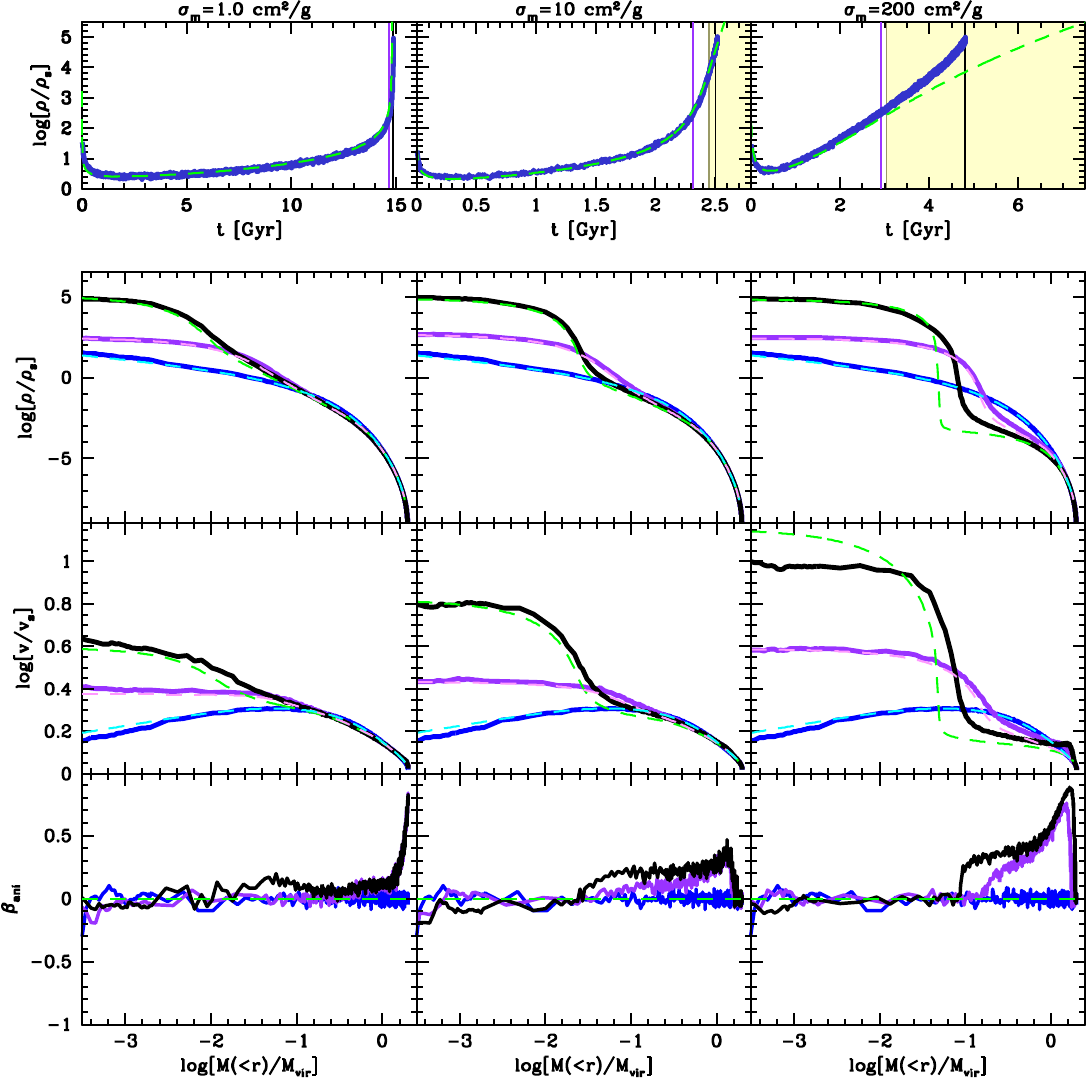}
  \caption{A detailed comparison of radial profiles obtained using the \SphMC code (solid curves), and those predicted using the (newly calibrated) \GTF code (dashed curves). Similar to Fig.~\ref{fig:profsNFW}, all results shown are for the NFWc30 halo and for three different values of $\sigma_\rmm$, as indicated at the top. The top panels show the evolution of the core density as a function of time. Similar to Fig.~\ref{fig:evolNFW} the yellow-shaded regions indicate the times when $r_{\rm fail} > r_\rmc$, which once again coincides with where the \GTF predictions deviate from the \SphMC simulations. The lower panels show the profiles of density, 1D velocity dispersion and velocity anisotropy parameter, plotted as a function of the enclosed mass. The \SphMC results correspond to the time indicated by the horizontal line of matching color in the top panels, while the \GTF results are for the time when the core density matches that of the \SphMC simulation results, which is not necessarily at the same time.  Clear discrepancies between the \GTF and \SphMC results are apparent at late times especially when $r_{\rm fail} > r_\rmc$. See text for a detailed discussion.}
\label{fig:profcomp}
\end{figure*}

Using a series of \SphMC simulations of this halo for different $\sigma_\rmm$, we obtain the core collapse times shown as red dots in Fig.~\ref{fig:t3ABG}. Using these results, we tune $\beta$ such that the \GTF code reproduces the results for $\tilde{\sigma}_\rmm \lta 0.1$. This yields $\beta=0.94$, which is slightly larger than for NFW haloes. The cyan curve shows the predictions obtained using the `standard' interpolation scheme (i.e., $\alpha=1$). Similarly to the NFW case, this significantly underestimates $t_3$ for $-1 \lta \log[\tilde{\sigma}_\rmm ] \lta 1.5$. The blue curve shows the predictions obtained with $\alpha=0.5$, which is the value that best fits the NFW data. Although this improves the agreement with the simulation results, the core collapse times for intermediate $\tilde{\sigma}_\rmm$ are now slightly over-estimated. We find that we can accurately fit the results for our \Core halo if we adopt $(\alpha,\beta) = (0.57, 0.94)$, indicated by the violet curves. Hence, unfortunately, but not unexpectedly, we infer that the values for $\alpha$ and $\beta$ need to be calibrated for each individual density profile, at least if the goal is to reach percent-level precision. 

\subsection{Unsolved issues regarding gravothermal evolution deep into the SMFP regime}
\label{sec:conundrum}

As is evident from the bottom panel of Fig.~\ref{fig:tcct0}, the \GTF model predicts collapse times that are somewhat larger than in the simulations, by up to $10$ percent. This discrepancy is also apparent in Fig.~\ref{fig:evolNFW}, which shows the evolution of core density $\rho_\rmc$ (left panels) and core velocity dispersion $v_\rmc$ (right panels), for the NFWc30 halo for four different values of $\sigma_\rmm$, as indicated. The azure-blue lines show the results obtained with the \SphMC code, while the black dashed lines indicate the results of the \GTF code with $(\alpha,\beta)=(0.5,0.9)$. For clarity, the time axis is normalised by the value of $t_3$ inferred from the \SphMC simulation. Overall, the agreement is very good. However, for large $\sigma_\rmm$ and at late times the \GTF code seems to predict somewhat lower densities and velocity dispersions than the \SphMC simulations. We have verified that these discrepancies are not manifestations of realisation noise (cf. Appendix~\ref{App:realisation}), and that they are robust to changes in the time stepping criteria used in the simulations (i.e., changes in $\eta$ or $\zeta$) or to changes in the spatial or temporal resolution of the \GTF code. We suspect instead that the discrepancy is due to the limited mass resolution of the \SphMC simulations. As discussed in Section~\ref{sec:spatiotemporal}, as the central density of a halo undergoing core collapse increases, there comes a point in time when the \SphMC code can no longer resolve the mean-free path in the central region. As long as the spatial extent $r_{\rm fail}$, over which this is the case is small compared to the core radius, this should not have a major impact, given that the core is largely homogeneous and isotropic. However, if $r_{\rm fail}$ becomes larger than or comparable to the core radius, the core will transport too much of its heat outward, thereby accelerating its collapse. The yellow shaded regions indicate the times when $r_{\rm fail} > r_\rmc$. This roughly coincides with the times when the densities and velocity dispersions of the \SphMC simulations exceed the \GTF predictions, supporting our hypothesis

However, there is also an alternative explanation for the discrepancy, first pointed out by \citet{Gurian.May.25}. Due to the development of velocity anisotropy in the simulations, something that is not captured by the \GTF code, heat transfer is no longer restricted to conduction, but can also arise from viscous dissipation of anisotropic stress (simply put, the stress tensor is no longer diagonal). In this case, there is excess heat transfer, not due to a lack of spatial resolution, but due to a real physical process that is accounted for in the \SphMC simulations but not in the \GTF code. Fig.~\ref{fig:profcomp} compares the radial profiles of the evolving NFWc30 halo obtained using the \SphMC code (solid lines) and the newly calibrated \GTF code (dashed lines). Similarly to Fig.~\ref{fig:profsNFW}, the results are shown for three different collision cross sections, as indicated, and the top panels show the temporal evolution of the core density. As in Fig.~\ref{fig:evolNFW}, the yellow shaded regions indicate the times when $r_{\rm fail} > r_\rmc$. The bottom panels show the profiles of density, 1D velocity dispersion, and the anisotropy parameter as a function of the enclosed mass. Each panel shows three profiles; the initial profile at $t=0$ (blue and cyan-dashed), an intermediate profile corresponding to $\rho_\rmc \sim 300 \rho_\rms$ (violet and pink-dashed), and a final profile corresponding to $\rho_\rmc = 10,000 \rho_\rms$ (black and green-dashed). In most cases, the profiles obtained using the \GTF code (dashed curves) are in good agreement with those from the \SphMC simulation. However, the final profiles for $\sigma_\rmm = 200 \cmg$ reveal large discrepancies; the core-within-a-core structure is clearly more pronounced in the \GTF density profile than in that inferred from the \SphMC simulation. The inner \GTF core is also hotter and smaller. Similar, albeit smaller, discrepancies are also apparent in the final profiles for $\sigma_\rmm = 1.0\cmg$ and $\sigma_\rmm = 10\cmg$ \citep[see also][]{Gurian.May.25}. If these discrepancies are due to viscous heat transfer that is not accounted for in the \GTF code, we expect to see a significant velocity anisotropy where the discrepancies are apparent. Upon close inspection of Fig.~\ref{fig:profcomp}, it is evident that the velocity distributions in the \SphMC simulations seem to become radially anisotropic at just about the region where the discrepancy arises. Although this supports the hypothesis that the discrepancies arise as a consequence of viscous dissipation of anisotropic stress, the alternative explanation, that they arise as a consequence of inadequate spatial resolution in the \SphMC simulations, cannot be ruled out. After all, as is apparent from the top panels, the discrepancies seem to be present only when $r_{\rm fail} > r_\rmc$.

\citet{Gurian.May.25} tested the viscosity-hypothesis by explicitly demonstrating that the heat flow in their {\tt KiSS-SIDM} simulation (similar to the \SphMC simulations used here) is inconsistent with that expected according to the gravothermal fluid equations. They then argue that "this deviation is naturally explained by viscous dissipation sourced in the intermediate mean-free-path regime". While true, it can also be explained by heat transfer that is artificially too large in the {\tt KiSS-SIDM} simulation due to insufficient spatial resolution.

Hence, we conclude that it is currently unclear whether the discrepancies shown here and in  \citet{Gurian.May.25} arise from artificially enhanced heat transport due to a lack of spatial resolution in the \SphMC simulations or from viscous heat transport that is not accounted for in the \GTF code. In principle, one could discriminate between these two alternatives by comparing simulations of different mass resolutions. If the lack of spatial resolution is the main cause of the discrepancy, simulations with smaller (larger) $N_\rmp$, should start to deviate from the \GTF predictions at earlier (later) times. If, instead, the discrepancy is due to viscous dissipation, it should be largely independent of $N_\rmp$. Unfortunately, this test is hampered by the fact that for $N_\rmp < 10^6$ realisation noise becomes too large (see Appendix~\ref{App:realisation}), while simulations with $N_\rmp > 10^6$ quickly become too CPU intensive. 

\section{The Core Collapse time of NFW haloes}
\label{sec:tccNFW}

\begin{figure*}
\centering
\includegraphics[width=.98\textwidth]{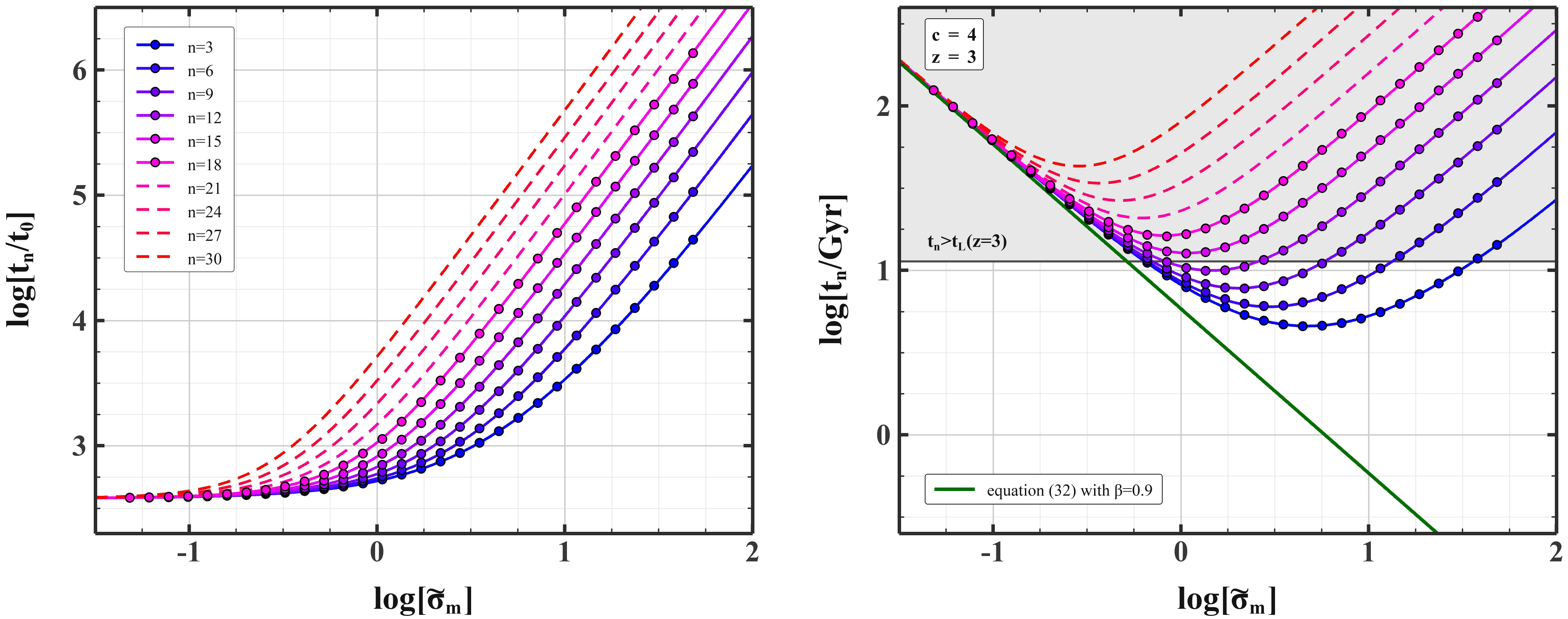}
  \caption{{\it Left panel:} The ratio $t_n/t_0$ for NFW haloes of arbitrary concentration, for different values of $n$ (different colours), as indicated, plotted as a function of the dimensionless cross section $\tilde{\sigma}_\rmm$. Here $t_n$ is defined as the time when the core density reaches $\log[\rho_\rmc/\rho_\rms]=n$. Filled circles are results obtained using the \GTF code, while coloured lines indicate the fitting function given by equations~(\ref{tnfitNFW})-(\ref{Akn}). We use dashed lines for values of $n$ for which we extrapolate the fitting function into a regime that is not probed by the \GTF calculations. {\it Right panel:} The time $t_n$ in Gyr as a function of $\tilde{\sigma}_\rmm$ for a NFW halo that virializes with a concentration parameter $c=4$ at redshift $z=3$. The grey-shaded region indicates where $t_n$ exceeds the lookback time, $t_L$, to $z=3$, while the green line indicates the core-collapse time given by equation~(\ref{tccEssig}), which is often used, incorrectly, to model the time-scale on which a SIDM halo collapses to form a BH.} 
\label{fig:tnsigma}
\end{figure*}

Using our newly calibrated \GTF model, we provide a simple recipe for computing the core collapse time of an NFW halo. We first use the \GTF model to follow the gravothermal collapse of the NFWc30 halo until its central core density reaches $\rho_\rmc = 10^{18} \rho_\rms$. In order to have adequate spatial resolution to resolve the core-within-a-core structure that forms during the late stages of core collapse we use $N=4000$ Lagrangian cells sampling the initial radial range from $r_{\rm min} = 5 \times 10^{-4} r_\rms$ to $r_{\rm max} = 200 r_\rms$. We use these results to measure $t_n$, which we define as the time it takes the halo to core collapse to a core density of $\rho_\rmc = 10^n \rho_\rms$. The filled circles in the left-hand panel of Fig.~\ref{fig:tnsigma} show the values of $t_n/t_0$ thus obtained. These are valid for NFW haloes of any concentration, and for any truncation that satisfies $M_{\rm tot} \gta 2\Mvir$. Note that $t_0$ is the collisional relaxation time of equation~(\ref{tcoll}), and not $t_n$ for $n=0$. Different colours correspond to different values of $n$ ranging from $3$ to $18$, as indicated. For $\tilde{\sigma}_\rmm \lta 0.1$ the results are virtually independent of $n$, which reflects the steep, super-exponential increase in density during core collapse in the LMFP regime. For larger values of $\tilde{\sigma}_\rmm$, the `collapse time' $t_n$ increases with $n$, reflecting the much slower collapse in the SMFP regime. 

These results are accurately described by the following fitting function
\begin{equation}\label{tnfitNFW}
{t_n \over t_0} = a_0 \, 
\left[ 1 + \sum_{k=1}^{N_k} A_k(n) \, \tilde{\sigma}_\rmm^{p_k} \right]\,.
\end{equation}
which is valid for $n\geq 3$. Here, 
\begin{equation}\label{Akn}
A_k(n) = \exp\left[a_k + b_k\,u + c_k\,u^2 \right]
\end{equation}
with $u = \ln(n/3)$. We require that $p_k>0$, such that each term in the sum vanishes in the limit $\tilde{\sigma}_\rmm \to 0$, which ensures that $(t_n/t_0) \to a_0$. In addition, the fit is performed under the physical monotonicity constraint $b_k + 2 \, c_k \, u > 0$, which ensures that $t_n$ always increases with $n$. We find that $N_k=3$ (a total of 13 free parameters) yields an accurate fit to the 180 \GTF data points with an RMS fractional error of $0.7\%$ and a maximum fractional error of $1.4\%$. Larger values for $N_k$ are not warranted by the data, while the quality of the fit is significantly worse for $N_k=2$. 
\renewcommand{\arraystretch}{1.05}
\begin{table}
\label{tab:fitpars}
\centering
\begin{tabular}{c|rrrr}
\hline
k &   $a_k$    &    $b_k$  &    $c_k$  & $p_k$\\
\hline
\hline
0 &  $380.648$ &      --   &      --   &  --  \\
1 & $-1.07694$ & $0.00595$ & $0.00111$ & $1.07645$ \\
2 & $-3.93730$ & $1.67739$ & $0.08505$ & $2.03409$ \\
3 & $-4.06945$ & $0.24146$ & $1.11046$ & $2.00711$  \\
\hline
\end{tabular}
\caption{Best-fit parameters of the fitting function specified by equations~(\ref{tnfitNFW})-(\ref{Akn})
describing the core collapse of NFW haloes of arbitrary concentration and for any truncation for which the total mass $M_{\rm tot} \gta 2 \Mvir$. Note that $b_k$, $c_k$ and $p_k$ are not defined for $k=0$.}
\end{table}

The solid lines in Fig.~\ref{fig:tnsigma} show the predictions obtained using equation~(\ref{tnfitNFW}) with the best-fit parameters listed in Table~2. The dashed lines show the predictions for $n=24$, $27$ and $30$ obtained by extrapolating the fitting function to larger $n$. 

Upon inspection, it is clear that $t_n$ scales as $\tilde{\sigma}_\rmm^{-1}$ in the limit of small cross sections and as $\tilde{\sigma}^p_\rmm$, with $p \sim 1$ in the limit of large $\tilde{\sigma}_\rmm$. This highlights an important point; the time-scale for core collapse does {\it not} monotonically decline with increasing cross section \citep[see also][]{Balberg.etal.02, Koda.Shapiro.11}. Rather, for given $n$ it reaches a minimum for a cross section $\tilde{\sigma}_\rmm$ that satisfies
\begin{equation}\label{sigmamin}
\sum_{k=1}^{N_k} A_k(n) \, (p_k-1) \, \tilde{\sigma}_\rmm^{p_k} = 1\,. 
\end{equation}
Unfortunately, there are no closed-form expressions for either $\tilde{\sigma}_\rmm^{\rm min}(n)$ or the corresponding $t_n^{\rm min}$, but both are trivially computed numerically. For $n=3$ this yields $t_3^{\rm min} = 131 \,t_{\rm core}$, which is reached for $\tilde{\sigma}_\rmm = 4.721$. Hence, an NFW halo takes at least $131$ core-crossing times to undergo core collapse to a central core density that exceeds a thousand times its original characteristic density. Note that $t_n^{\rm min}$ is a strong function of $n$. For example, the minimum time  to reach a core density $\rho_\rmc = 10^{9}\rho_\rms$ ($10^{15}\rho_\rms$) is $221\, t_{\rm core}$ ($362\, t_{\rm core}$) and requires a cross section $\tilde{\sigma}_\rmm = 2.023$ ($1.085)$. 

Using the definition of the core-crossing time, we can rewrite equation~(\ref{tnfitNFW}) as
\begin{equation}\label{tnfitGyr}
t_n = 165.0 \, h^{-1} \Gyr \, {g(c) \over \tilde{\sigma}_\rmm \, \Delta_{200}^{1/2} \, E(z)} \,
\left[ 1 + \sum_{k=1}^{N_k} A_k(n) \, \tilde{\sigma}_\rmm^{p_k} \right] \,.
\end{equation}
Here, $g(c) = \sqrt{f(c)/c^3}$, $\Delta_{200} = \Delta_{\rm vir}/200$ and $E(z) = H(z)/H_0$ with $H(z)$ the Hubble parameter. As an illustration, the right-panel of Fig.~\ref{fig:tnsigma} shows this $t_n$ as a function of $\tilde{\sigma}_\rmm$ for an NFW halo that virializes at redshift $z=3$. We have adopted $\Delta_{\rm vir}=200$ and assumed a concentration $c=4$, which is a typical value for a halo shortly after virialization \citep[][]{Wechsler.etal.02, Zhao.etal.03a, Zhao.etal.03b}. Different colours correspond to different values of $n$, following the convention of the left panel. The grey-shaded region indicates where $t_n$ is larger than the lookback time $t_\rmL(z)=11.36 \Gyr$ to $z=3$. What is apparent from this figure is that an NFW halo that forms at $z=3$ will not have enough time to collapse to a central density $\rho_\rmc > 10^{13} \rho_\rms$ at present, no matter the value of $\sigma_\rmm$. Reaching core densities $\rho_\rmc > 1000 \rho_\rms$ by $z=0$ requires $0.64 \lta \tilde{\sigma}_\rmm \lta 35.3$. 

\begin{figure*}
\centering
\includegraphics[width=.9\textwidth]{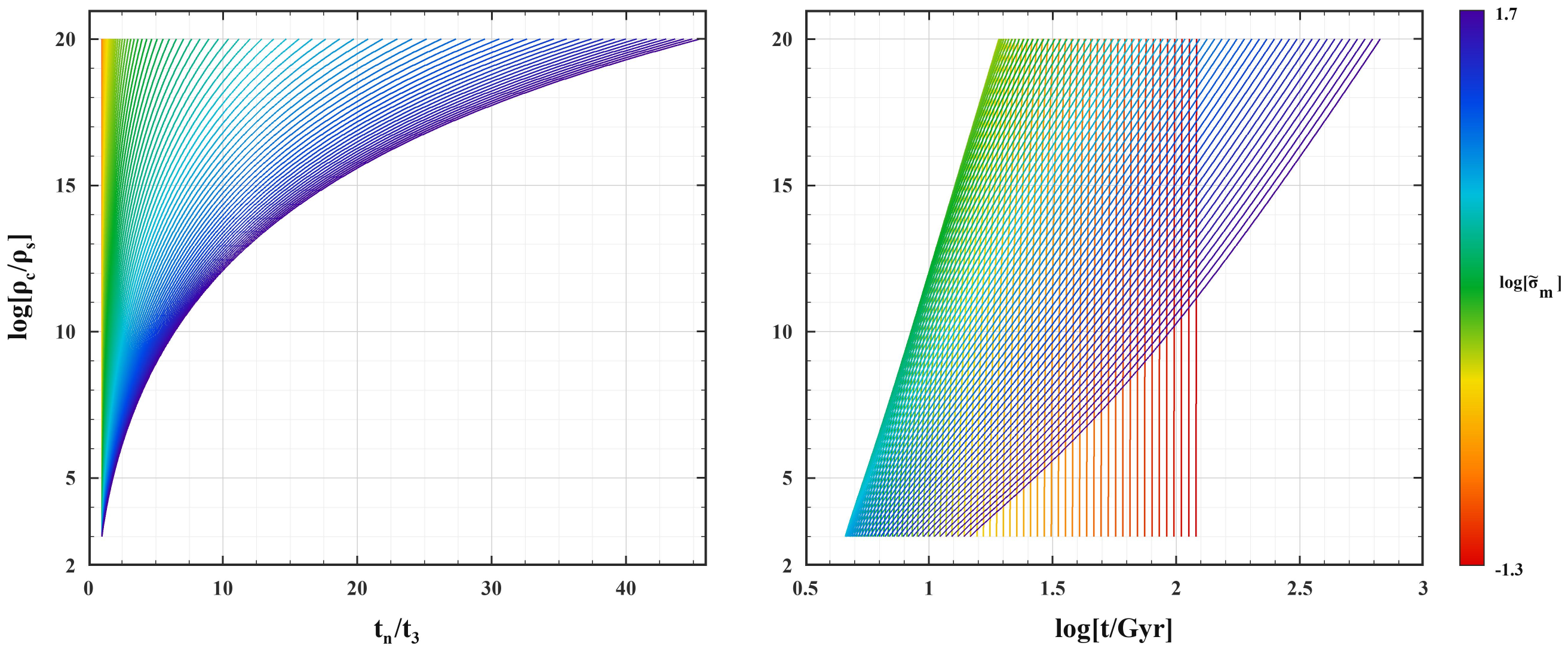}
  \caption{Evolution of the central density of an NFW halo $\rho_\rmc$, in units of the characteristic density $\rho_\rms$, undergoing gravothermal core collapse as inferred from the fitting function discussed in Section~\ref{sec:tccNFW}. Different colored lines correspond to different values of the normalized collision cross section, $\tilde{\sigma}_\rmm$, as indicated by the color-bar shown to the right.  Results are shown both as a function of $t_n/t_3$ (left-panel) and as a function of time in Gyr (right-panel). Whereas the former are valid for {\it any} NFW halo, independent of its concentration, the explicit times in the right-hand panel are computed for a NFWc30 halo that virializes with a concentration parameter $c=4$ at $z=3$. Note that larger cross sections clearly result in a reduced rate of core collapse, when normalized by $t_3$. When expressed in Gyr, there is a minimum time needed to collapse to a certain core density; the minimum time to reach a higher central density requires a smaller $\tilde{\sigma}_\rmm$.}
\label{fig:rhocfit}
\end{figure*}

Using the fitting function provided above, it is straightforward to compute any $t_n$ for any NFW halo of arbitrary concentration, at any redshift, and for any (integer or non-integer) value of $n\geq 3$. Hence, it can be used to construct the full temporal evolution of the core density $\rho_\rmc$ of any NFW halo. This is illustrated in Fig.~\ref{fig:rhocfit}, which depicts $\rho_\rmc$ as a function of $t/t_3$ (left-hand panel) and as a function of time in Gyr (right-hand panel), computed directly from equations~(\ref{tnfitNFW})-(\ref{Akn}). Results are shown for 100 different values of $\tilde{\sigma}_\rmm$ logarithmically sampled over the range from $0.05$ to $50$, as indicated. Note that the curves start at $\rho_\rmc = 10^3 \rho_\rms$, as the fitting function only describes the behaviour past this point. The left panel clearly demonstrates the reduced rate of core collapse for larger cross sections. The right panel, which is computed assuming an NFWc30 halo with concentration $c=4$ at $z=3$, nicely depicts the presence of a minimum time needed to collapse to a certain core density, and that this minimum is achieved for a smaller $\tilde{\sigma}_\rmm$ for larger $\rho_\rmc$.

\subsection{Forming seed black holes with SIDM core collapse}
\label{sec:BH}

If nothing can stop the gravothermal catastrophe, the consensus is that the core will ultimately become relativistic, triggering an instability that causes the direct collapse to a BH. In what follows, we define $t_{\rm BH}$ as the time it takes for a halo's core velocity dispersion, $v_\rmc$, to become equal to $c/3$, with $c$ the speed of light, which we assume to coincide with the time of BH formation.

In the LMFP regime, the core density rises at a super-exponential rate. It is therefore tempting, but incorrect (see below), to assume that the core density will continue to increase at this same rate until the core becomes relativistic and collapses to a black hole. In this context, \citet{Pollack.etal.15} reported a time-scale for catastrophic collapse to a black hole of $t_{\rm BH} = 455.65\,t_0$ for a fiducial isolated NFW halo. Based on the self-similarity of gravothermal collapse in the LMFP regime, \citet{Essig.etal.19} derived a core-collapse time of
\begin{equation}\label{tccEssig}
t_{\rm cc} = {150 \over \beta} \, {1 \over \sigma_\rmm \, \rho_\rms \, r_\rms} \, {1 \over \sqrt{4 \pi G \rho_\rms}}\,.
\end{equation}
which can be rewritten as
\begin{equation}\label{tccEssigalt}
t_{\rm cc} = {150 \over \beta} \, {1 \over \tilde{\sigma}_\rmm} \, t_{\rm core} = {338.5 \over \beta} \, t_0\,.
\end{equation}
Using $\beta=0.75$, which is a typical value that is often quoted, this translates to $t_{\rm cc} = 451.35 t_0$, very similar to the value obtained by \citet{Pollack.etal.15}. Using $\beta=0.9$, which is the value advocated here, this yields $t_{\rm cc} = 376.1 t_0$ which is a good match to equation~(\ref{tnfitNFW}), but only for sufficiently small $\tilde{\sigma}_\rmm$. 

In recent years, numerous studies have used equation~(\ref{tccEssig}) to model the time-scale for core collapse and have assumed that this coincides with the time-scale on which core collapse produces a seed BH \citep[][]{Choquette.etal.19, Shah.Adhikari.24, Roberts.etal.25b, Roberts.etal.25c, Jiang.etal.26}. However, as clearly stressed by \citet{Balberg.etal.02}, \citet{Essig.etal.19} and others, and as evident from Fig.~\ref{fig:tnsigma}, this self-similar scaling is only valid in the LMFP regime, an important constraint that seems to have been overlooked on more than one occasion. A proper derivation of $t_{\rm BH}$ needs to account for the fact that {\it any} collapsing core becomes optically thick (and thus enters the SMFP regime) before it becomes relativistic. And when the core is no longer in the LMFP regime, its density and velocity dispersion increase much slower. As a consequence, the core collapse time given by equation~(\ref{tccEssig}), which is only valid in the LMFP regime, can severely underestimate $t_{\rm BH}$. 

To illustrate this point, the solid green line in the right-hand panel of Fig.~\ref{fig:tnsigma} indicates $t_{\rm cc}$ obtained using equation~(\ref{tccEssig}) with $\beta=0.90$. This accurately fits the results of our \GTF calculations, but only for $\tilde{\sigma}_\rmm \lta 0.1$. For larger cross sections, equation~(\ref{tccEssig}) drastically underestimates $t_n$, for any $n$. Relating this to the time-scale for BH formation requires computing the core density at the time the velocity dispersion of the core becomes relativistic (i.e., when $v_\rmc \simeq c/3$), which we address in detail in a follow-up paper (Tokayer \& van den Bosch, in prep.). Here we present a rough estimate based on \citet{GadNasr.etal.24}, who, using a GTF code, find that during the late stages of gravothermal collapse, when the core is well in the SMFP regime, the density and velocity dispersion of the core evolve according to $\rmd\log \rho_\rmc/\rmd\log v_\rmc \sim 10$. Let $\rho_n = 10^n \rho_\rms$ and $v_n$ be the density and velocity dispersion of the core when it reaches this stage. Then, the core density at the moment the core becomes relativistic is given by
\begin{equation}\label{rhocc}
\log\left({\rho_\rmc \over \rho_\rms}\right) =  n + {\rmd\log\rho_\rmc \over \rmd\log v_\rmc} \log\left({c/3 \over v_n}\right) 
\end{equation}
As an example, consider the NFWc30 halo with a collision cross section of $\sigma_\rmm = 200 \cmg$. As is evident from the right-hand panels of Fig.~\ref{fig:evolNFW}, when this system reaches a core density $\rho_\rmc = 10^5 \rho_\rms$, the velocity dispersion of the core has reached $v_\rmc \approx v_\rms = 566.41\kms$. Since $\tilde{\sigma}_\rmm = 32$, it is well into the SMFP regime at this point in time, and we may therefore assume that the core will continue to evolve obeying $\rmd\log \rho_\rmc/\rmd\log v_\rmc \sim 10$. Using equation~(\ref{rhocc}), we infer that $\log(\rho_\rmc/\rho_\rms) \approx 27.5$. Thus, we can estimate the time for BH formation as $t_{\rm BH} = t_n$ with $n=27.5$. Using the fitting function given by equations~(\ref{tnfitNFW})-(\ref{Akn}), this yields $t_{\rm BH} = 3.26 \times 10^6 t_0$, orders of magnitude larger than simply equating $t_{\rm BH}$ with $t_{\rm cc}$ as in numerous previous studies. In particular, although the collisional relaxation time for the NFWc30 halo listed in Table~1 is only $t_0 = 0.186 \Myr$ (at $z=0$), it will take a staggering $604.9 \Gyr$ for this halo to collapse to form a seed BH. 

Recently, \citet{Gu.etal.26} presented a new method for solving the gravothermal fluid equations that allows them to follow the core collapse of SIDM haloes to the point of BH formation by incorporating general relativistic effects through the  Misner-Sharp formalism. They model the conductivity using the standard interpolation scheme ($\alpha=1.0$) with $\beta=0.75$, and apply their formalism to an NFW halo with $\rho_\rms = 0.019\Msun\pc^{-3}$, $r_\rms = 2.6\kpc$ and $v_\rms = 83.6 \kms$, and using a collision cross section $\sigma_\rmm = 5 \cmg$ which implies $\tilde{\sigma}_\rmm = 0.052$. As expected, the system starts in the LMFP regime with a phase of super-exponential core collapse at $t \approx 450 t_0$. However, once the core becomes optically thick, the rate at which the core density increases slows down significantly. Using a multi-stage approach, in which the authors repeatedly zoom in to smaller and smaller scales, they are able to follow the collapse all the way up to horizon crossing, which occurs at $t_{\rm BH} = 580 t_0$. This confirms that equation~(\ref{tccEssig}) underestimates the time-scale for BH formation, since it does not account for the fact that at some point the core always becomes optically thick. Since $\tilde{\sigma}_\rmm$ is very small, this happens at a relatively late stage, when $\rho_\rmc$ is already very large. This explains why $t_{\rm BH}/t_{\rm cc} \approx 580/450 = 1.29$ is fairly small. This is consistent with our results, which show that $t_n/t_{\rm cc} \to 1$ in the limit $\tilde{\sigma}_\rmm \to 0$. Note, though, that the collisional relaxation time for the halo studied in \citet{Gu.etal.26} is $t_0 = 0.26 \Gyr$, Hence, the time-scale for BH formation is $t_{\rm BH} = 151 \Gyr$, clearly too long to be relevant. 

The characteristic radius and density of the halo used by \citet{Gu.etal.26} are consistent with a halo that virializes at $z=2$ with a virial mass $\Mvir = 6.3 \times 10^9\Msun$ and a concentration parameter $c=10$ (assuming $\Delta_{\rm vir} = 97$). To reasonable approximation, $r_\rms \propto (1+z)^{-1}$, $\rho_\rms \propto (1+z)^{3}$ and $v_\rms \propto (1+z)^{1/2}$. If we keep $\sigma_\rmm = 5 \cmg$, then the collisional relaxation time $t_0 = (a \rho_\rms \sigma_\rmm v_\rms)^{-1} \propto (1+z)^{-7/2}$. Hence, assuming that the halo virializes at a higher redshift, we can significantly reduce $t_0$. For example, for  $z=15$ we have that $t_0$ is only $0.74\Myr$. If we once again assume that $t_{\rm BH} = 580 t_0$, this would imply that BH formation only takes $\sim 430 \Myr$. However, this is incorrect. After all, $\tilde{\sigma}_\rmm = \sigma_\rmm \rho_\rms r_\rms \propto (1+z)^2$. Hence, when $z$ increases from 2 to 15, while keeping $\sigma_\rmm$ fixed, $\tilde{\sigma}_\rmm$ increases from $0.052$ to $1.48$. As is evident from Fig.~\ref{fig:tnsigma}, at such a large $\tilde{\sigma}_\rmm$ the ratio $t_{\rm BH}/t_0$ is expected to be orders of magnitude larger. For example, using our model, we find that for $\tilde{\sigma}_\rmm = 1.48$ it takes $\sim 1800 t_0$ ($10,800 t_0$) to reach a core density $\rho_\rmc = 10^{18}\rho_\rms$ ($10^{30} \rho_\rms$). For comparison, the core of the halo modelled in \citet{Gu.etal.26} reaches a density $\rho_\rmc \approx 10^{35} \rho_\rms$ at the time of BH formation\footnote{This is inferred from the bottom panel of their Fig.~5, according to which the core at BH formation has a mass $M_\rmc \sim 3\times 10^{-8} \Mvir \sim 200 \Msun$ and a radius $r_\rmc = 10^{-14} r_\rms  = 2.6 \times 10^{-11} \pc$.} Therefore, we conclude that, in reality, the formation of the seed BH will take at least $(10,800/580) \times 0.43 = 8.0 \Gyr$. Starting from $z=15$, this would imply that the BH does not form until $z \simeq 0.5$ at the earliest. 

Alternatively, rather than keeping $\sigma_\rmm$ fixed at $5 \cmg$, one might consider reducing the physical collision cross section such that $\tilde{\sigma}_\rmm$ remains fixed at $0.052$. This ensures that the result of \citet{Gu.etal.26}, that $t_{\rm BH} = 580 t_0$, remains valid. However, this reduces $\sigma_\rmm$ from $5 \cmg$ to $0.176 \cmg$, which increases $t_0$ from $0.74\Myr$ to $21.0 \Myr$. Consequently, $t_{\rm BH} = 12.2 \Gyr$, which implies that the BH will not form until $z \approx 0.08$.

Although these estimates are crude at best, they make it clear that the formation of seed BHs via the gravothermal collapse of SIDM haloes takes much longer than previously assumed. In fact, taken at face value, these examples suggest that it is unlikely that this mechanism can explain the population of surprisingly massive BHs observed at high redshifts with JWST. In a follow-up paper (Tokayer \& van den Bosch, in prep) we address this issue in much greater detail.

\section{Conclusions}
\label{sec:concl}

We have studied the gravothermal evolution of SIDM haloes using a new spherical Direct Simulation Monte Carlo code called \SphMC that is similar to the kinetic solver {\tt KiSS-SIDM} recently developed by \citet{Gurian.May.25} and to the {\tt NSphere} code developed independently by \citet{Kamionkowski.Sigurdson.25a}. \SphMC assumes spherical symmetry and integrates the radial coordinates of particles in an $N$-body system, while treating collisional scattering using the Direct Simulation Monte Carlo (DSMC) method of \citet[][]{Bird.94}. The main advantages of this method are (i) that it is much faster than the $N$-body method, (ii) that it does not require any particle softening, (iii) that it allows integration to much higher central densities, and (iv) that it typically results in much better energy conservation than standard $N$-body codes. The main shortcoming of \SphMC is that it only works for spherically symmetric systems.

We have compared the results obtained with \SphMC with those obtained using the standard $N$-body method. The results of the latter converge to those obtained with \SphMC for small softening lengths, significantly smaller (roughly by a factor of four) than what is considered ideal for CDM simulations, in the sense that running the same system without self-interactions causes artificial core formation due to inadequate suppression of two-body interactions. We have also demonstrated (Appendix~\ref{App:Nbodysims}) that multiple scattering events can be allowed per time step in $N$-body simulations, despite many claims in the literature to the contrary. 

A comparison of \SphMC with results obtained using a \GTF code that solves the gravothermal fluid equations shows that the standard harmonic scheme used to interpolate the conductivity between the long- and short-mean-free paths is inaccurate. Allowing for a more gentle transition between these two regimes yields results that are in perfect agreement with the \SphMC results. In particular, we show that a single value for the parameter that controls the conductivity in the LMFP regime can reproduce the results for any NFW halo, independent of halo concentration or the extent to which the halo has been truncated. The fact that all studies to date have adopted an inaccurate interpolation function explains why there is a large discrepancy in the literature as to the value of this free parameter. The best-fit parameters that quantify the conductivity $\kappa(\sigma_\rmm)$ show a weak dependence on the detailed density profile of the halo in question. Hence, it is recommended that these parameters be calibrated against simulation results for density profiles other than those considered here.

For sufficiently small collision cross sections ($\tilde{\sigma}_\rmm \lta 0.1$), the early stages of core collapse take place in the LMFP regime in which the Knudsen number of the core remains above unity. During this stage, the density and effective temperature of the core increase super-exponentially. With our revised conductivity calibration, we find the time-scale for this catastropic core collapse phase to be $t_{\rm cc} = 380 \, t_0$, roughly 16 percent shorter than the value advocated in previous studies \citep[$\sim 455 \, t_0$; see e.g.,][]{Pollack.etal.15, Essig.etal.19, Jiang.etal.26}. Numerous studies in recent years have assumed that this catastropic collapse continues all the way to the point at which the core becomes relativistic and collapses directly, via a radial instability, to a black hole. However, this is incorrect. As first pointed out in \citet{Balberg.etal.02}, it is unavoidable for the core to become optically thick before it becomes relativistic. The evolution of the core during this SMFP regime is very different from that in the LMFP regime. As the core transitions from the LMFP to the SMFP regime, the mean free path in the core becomes smaller than the core radius, and the core can only transfer heat to the outskirts from its outer layers, rather than from its entire volume. This drastically slows the rate at which the core density and temperature evolve. In addition, a core-within-a-core structure develops, with the inner core continuing to undergo core collapse transferring its heat to the outer core, which expands and slowly transfers the heat to the halo outskirts. 

Following the gravothermal evolution during these later stages of core collapse is extremely challenging. Standard $N$-body simulations can barely follow the evolution of the core past the point where the core density reaches $\rho_\rmc \simeq 10^3 \rho_\rms$, as the time steps required become too small. The \SphMC code developed here can follow the collapse significantly further, to core densities that are several orders of magnitude larger. However, because of its finite spatial resolution (set by the number of particles used), at some point it becomes impossible to resolve the ever-shrinking mean-free path in the collapsing core. Failure to resolve $\lambda_{\rm mfp}$ causes artificially enhanced heat transfer, which in turn causes core collapse to proceed too rapidly. Indeed, we find that the \SphMC results start to deviate from those obtained using the \GTF code once the mean-free path is no longer resolved across the entire core. Unfortunately, \GTF codes also face challenges in resolving the core collapse deep into the SMFP regime. In particular, because they are inherently one-dimensional, they fail to account for the development of strong radial anisotropy just outside of the (inner) core. As pointed out in \citet{Gurian.May.25}, such anisotropy can give rise to viscous heat transport. Since this is not taken into account in the \GTF codes, the latter may {\it underestimate} the rate of heat transfer, causing core collapse to proceed too slowly. Currently, it is unclear whether the discrepancy between the \GTF and \SphMC results in the SMFP regime (depicted in Figs.~\ref{fig:evolNFW} and~\ref{fig:profcomp}) indicates a failure of the former, the latter, or both. Until this conundrum is resolved, the detailed gravothermal evolution deep into the SMFP regime remains unclear (see also Tokayer \& van den Bosch, in prep.).

Despite these shortcomings, we have used the \GTF code to follow the core collapse of NFW haloes all the way to core densities of $\rho_\rmc = 10^{18} \rho_\rms$. We used these results to construct an easy-to-use fitting function that describes the core collapse of NFW haloes (as inferred from the gravothermal fluid equations). Taking these results at face value, two important points can be made. First, the time $t_n$ for the core to collapse to a density $\rho_\rmc = 10^{n} \rho_\rms$ does not decline monotonically with $\sigma_\rmm$. Rather, it transitions from $t_n \propto \sigma^{-1}_\rmm$ for small cross sections to $t_n \propto \sigma^{p}_\rmm$ with $p \sim 1$ for large cross sections. The cross section for which $t_n(\sigma_\rmm)$ is minimal decreases with increasing $n$.  Secondly, previous estimates for the formation time $t_{\rm BH}$ of a seed black hole through the gravothermal core collapse of a SIDM halo have been drastically underestimated. In particular, it is typically assumed that $t_{\rm BH} = t_{\rm cc} \simeq 455 t_0$ \citep[][]{Pollack.etal.15, Choquette.etal.19, Shah.Adhikari.24, Roberts.etal.25b, Roberts.etal.25c, Jiang.etal.26}. However, this is only valid in the LMFP regime and ignores that any core must, at some point, become optically thick before becoming relativistic, which slows down subsequent evolution \citep[see e.g.,][]{Balberg.etal.02, GadNasr.etal.24, Gu.etal.26}. In addition, it incorrectly assumes that $t_{\rm BH} \propto \sigma^{-1}_\rmm$, and can thus be made arbitrarily small by simply adopting a sufficiently large collision cross section. Based on the (admittedly crude) results presented here, we instead argue that forming seed BHs through the gravothermal collapse of a SIDM halo takes too long for this process to be a viable explanation for the alledged population of supermassive black holes at high redshifts discovered with JWST.  Although core collapse can be accelerated by invoking a variety of baryonic processes \citep[][]{Elbert.etal.18, Sameie.etal.18, Feng.etal.21, Zhong.etal.23}, it is unlikely that this is sufficient to overcome the bottleneck associated with collapse in the SMFP regime.

\section*{Acknowledgements}

We are grateful to Annika Peter, James Gurian, Charlie Mace, Carton Zeng, and Fangzhou Jiang for useful discussions. This work has been supported by the US National Science Foundation through grant AST-2407063.

\section*{Data Availability}
 
The SphDSMC code and the simulation data  created as part of this project are available from the authors upon request.


\bibliographystyle{mnras}
\bibliography{references_vdb} 


\appendix
\numberwithin{figure}{section}
\numberwithin{table}{section}
\numberwithin{equation}{section}

\section{N-body simulations of self-interacting dark matter}
\label{App:Nbodysims}

To account for DM self-interactions in $N$-body simulations, we closely follow \citet[][hereafter V+12]{Vogelsberger.etal.12}. We assume that a particle $i$ can only scatter with one of its $k=32$ nearest neighbours and compute the expectation value for the number of scattering events of particle $i$ in time step $\Delta t$ according to
\begin{equation}\label{Nscat}
\langle N_{{\rm scat},i} \rangle = \frac{1}{2} \sum_j \sigma_m \, m_\rmp \, W(r_{ij}, h_i) \, |\bv_j - \bv_i| \, \Delta t\,.
\end{equation}
Here, the summation is over the $k$ nearest neighbours, $m_\rmp$ is the particle mass, $W(r,h)$ is a kernel function that describes the `extent' of the simulation particle in configuration space, and $h_i$ is the distance enclosing the $k$ nearest neighbours of particle $i$. The factor $1/2$ is needed to avoid double counting. The actual number of interactions for particle $i$ is drawn from a Poisson distribution with an expectation value equal to $\langle N_{{\rm scat},i} \rangle$. As detailed below, this is where our approach differs from that of V+12 and many others.
\begin{figure*}
\centering
\includegraphics[width=0.95\textwidth]{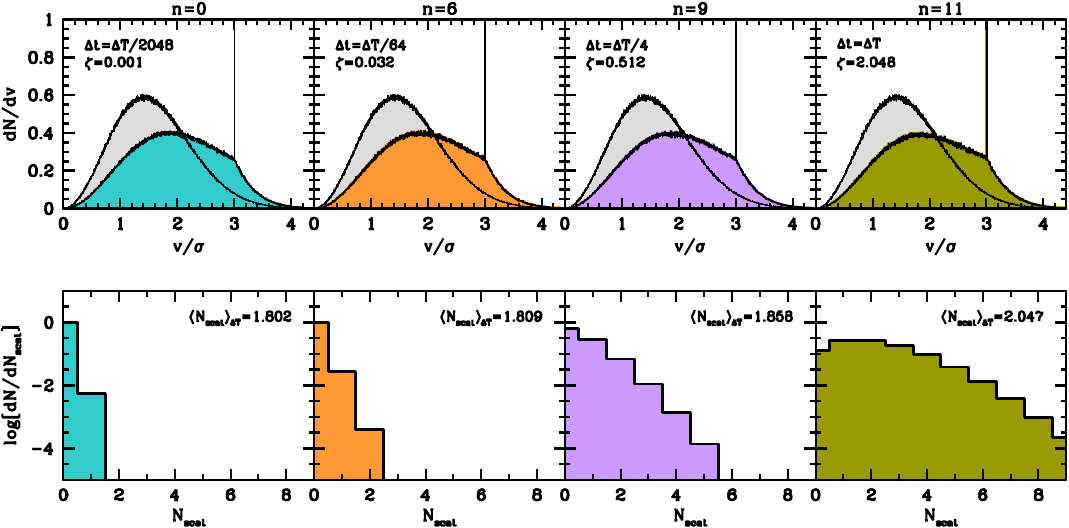}
  \caption{Results of Monte-Carlo simulations in which each of $5 \times 10^6$ particles, all with the same initial velocity $v_{\rm init} = 3\sigma$, are scattered against random particles from a heat bath of constant temperature $T = \sqrt{3}\sigma$. Each particle is integrated for a total duration of $\Delta T$, using $N_{\rm step}$ time steps of $\Delta t = \Delta T/N_{\rm step}$. Each time step $\Delta t$ the expectation value for the number of scatter events in the subsequent $\Delta t$ is updated based on the particle's new velocity using equations~(\ref{Nscatexp}) and~(\ref{vrelexp}). Different columns correspond to different $N_{\rm step}$, ranging from $2048$ in the left-most panel to $1$ in the right-most panel. Larger values of $N_{\rm step}$ correspond to a smaller effective value for $\zeta$, as indicated. {\it Top Panels:} Coloured histograms show the distributions of $v_\rmP/\sigma$ after a time $\Delta T$, while the grey-shaded histograms show the MB velocity distribution of the heat bath. The sharp peak at $v/\sigma=3$ corresponds to the particles that did not experience any scattering event during the entire time interval $\Delta T$. {\it Bottom Panels:} The (normalised) distributions of $N_{\rm scat}$, the number of scattering events particle $P$ experienced per time step $\Delta t$. Despite large differences in the latter, the PDFs of the final velocities for $P$ are virtually indistinguishable. This justifies the use of multiple scattering events per time step.}
\label{fig:multiscat}
\end{figure*}

We impose the following constraint on the time step
\begin{equation}\label{timestepSI}
 \Delta t < (\Delta t)_{\rm SI} \equiv \frac{\zeta}{{\rm max} (\frac{1}{2} \, \rho_i \, \sigma_\rmm \, \langle v_{\rm rel} \rangle_i)} \,.
\end{equation}
Here, $\zeta$ is a free parameter, $\rho_i = m_\rmp \sum_j W(r_{ij},h_i)$, $\langle v_{\rm rel} \rangle_i = {1 \over k} \sum_j |\bv_j - \bv_i|$ where the sum is over the $k$ nearest neighbours of particle $i$, and the maximum in the denominator is taken over all $N$ simulation particles $i$. It is common to pick a value for $\zeta$ such that the probability of a particle to scatter more than once per time step is negligible. For example, V+12 use $\zeta \sim 0.01$, while others have advocated even smaller values \citep[][]{Vogelsberger.etal.19, Mace.etal.24, Palubski.etal.24}.  

Our approach is different. We set $\zeta = 0.5$ and allow multiple scattering events per particle per time step. In particular, once it has been determined that a target particle $i$ has $N_{\rm scat}\geq 1$ scattering events in time step $\Delta t$, we proceed as follows. For each scattering event particle $i$ is scattered against one of its $k$ nearest neighbours, where the probability to scatter with particle $j$ is given by
\begin{equation}
P_{ij} = \frac{W(r_{ij}, h_i) \, |\bv_j - \bv_i|}{\sum_j W(r_{ij}, h_i) \, |\bv_j - \bv_i|}\,,
\end{equation}
Once a particle $j$ is selected for a collision with $i$, we update their velocities following the same procedure as for the \SphMC code (see Section~\ref{sec:SphMC}).

Throughout, we adopt the standard (normalised) cubic spline function
\begin{equation}\label{spline}
W(r,h) = \frac{8}{\pi h^3} \left\{ \begin{array}{ll}
1 - 6q^2 - 6 q^3 & 0 \leq q \leq \frac{1}{2} \\
2(1 - q)^2       & \frac{1}{2} < q \leq 1 \\
0                & q > 1
\end{array}
\right. \,,
\end{equation}
where $q=r/h$. However, we have also experimented with an alternative top-hat kernel which yields results that are very similar. 

All particles advance with the same time step $\Delta t$, whose value is adjusted each time step so that
\begin{equation}\label{timestep}
\Delta t = {\rm min}\left[ (\Delta t)_{\rm SI}, (\Delta t)_{\rm grav} \right]\,.
\end{equation}
Here $(\Delta t)_{\rm SI}$ is the time step for the self-interactions given by equation~(\ref{timestepSI}) and 
\begin{equation}\label{timestepgrav}
 (\Delta t)_{\rm grav} = \sqrt{2 \, \eta \, \frac{\epsilon}{{\rm max} |\ba_i|}}
\end{equation}
is the time step needed to adequately resolve the orbits due to the gravitational force from all other particles. Here $\ba_i$ is the gravitational acceleration of particle $i$, $\epsilon$ is the Plummer softening length, and $\eta$ is a free parameter which we set to $\eta = 0.025$ throughout. Although adjusting $\Delta t$ each time implies a loss of symplecticity of our integration scheme, we find that the total energy in our simulations is conserved to better than one part in $10^4$ up to the final stages of core collapse (see Fig.~\ref{fig:compNFWc40}).

\subsection{Multiple Scattering}
\label{sec:multiscat}

The key difference between our approach and that of V+12 is that we interpret the rhs of equation~(\ref{Nscat}) as an expectation value for the number of scatter events, while V+12 interpret it as the probability $P_1$ that the particle in question scatters once, and only once, in time step $\Delta t$. Given that the formal expression for $P_1$ is given by
\begin{equation}\label{Pone}
P_{1} = \langle N_{\rm scat} \rangle \, \exp^{-\langle N_{\rm scat} \rangle}
\end{equation}
we see that the two approaches become effectively equal in the limit where $P_1$ is very small. This in turn requires a sufficiently small time step, and thus a small value for $\zeta$.

To our knowledge, all numerical simulations of SIDM to date have adopted the same approach as V+12, to which we refer as the `standard method' in what follows. It is based on the notion that one should not allow a particle to scatter more than once per timestep. The motivation for this restriction is that once a particle has scattered, its velocity changes, which impacts its subsequent scatter probability. However, as we demonstrate below, allowing multiple scattering events per particle per time step is not a concern (within limits), as long as one properly interprets the rhs of equation~(\ref{Nscat}) as an expectation value. This means that one can allow values for $\zeta$ that are actually much larger than the small values usually adopted in the standard model, which can drastically speed up the simulation especially during the core collapse phase.

In order to probe the impact of multi-scattering per time step, we perform a simple idealised experiment. Consider a particle $P$ of initial velocity $v_{\rm init}$ in a sea of particles that follow a Maxwell-Boltmann (MB) velocity distribution with 1D velocity dispersion $\sigma$ (i.e., a thermal bath of temperature $T = \sqrt{3}\sigma$). We assume a homogeneous density of arbitrary value and a collisional cross section per unit mass, also of arbitrary value, and pick a time step $\Delta t$ such that $\langle N_{\rm scat} \rangle = 0.001 \times 2^n$. Here, $n=0,1,2,...,11$ is the one free parameter of our experiment and
\begin{equation}\label{Nscatexp}
\langle N_{\rm scat} \rangle = 0.5 \, \sigma_\rmm \, \rho \, \langle v_{\rm rel} \rangle \, \Delta t\,.
\end{equation}
The expectation value for the relative velocity of $P$ with respect to a random particle drawn from the MB distribution is analytical and given by
\begin{equation}\label{vrelexp}
\langle v_{\rm rel} \rangle =\sigma \left[ \sqrt{2 \over \pi} \exp\left(-{x^2\over 2}\right) + \left(x + {1 \over x}\right) \, {\rm erf}\left({x \over \sqrt{2}}\right)\right] \,,
\end{equation}
where $x = v_\rmP/\sigma$. For each value of $n$, we `evolve' $P$ for a total integration time of $\Delta T = N_{\rm step} \, \Delta t$ where $N^{\rm step} = 2^{11-n}$. Note that this leaves $\Delta T$ invariant among all experiments (i.e., for all $n$). Larger values of $n$ implies fewer, but longer time steps $\Delta t$, and a larger expectation value for the number of scatter events per time step. Since $\zeta$ is equal to the maximum value of $\langle N_{\rm scat} \rangle$ allowed per time step (cf. equation~[\ref{timestepSI}]), as we increase $n$ we effectively increase $\zeta$ from $0.001$ ($n=0$) to $2.048$ ($n=11$). Hence, these experiments probe the impact of varying $\zeta$ by more than three orders of magnitude. 
\begin{figure}
\centering
\includegraphics[width=0.45\textwidth]{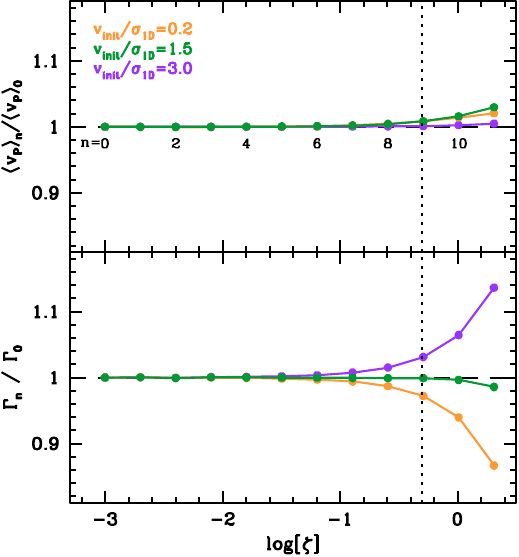}
  \caption{Summary statistics of the multi-scatter Monte-Carlo simulations. The top panel plots the ratio $\langle v_\rmP \rangle_\rmn / \langle v_\rmP \rangle_0$. Here, $\langle v_\rmP \rangle$ is the average velocity of particle $P$ after a fixed total integration time of $\Delta T$ and the subscript indicates the value of $n = 11-\log_{2}(\Delta T/\Delta t)$ used in the experiment, with $\Delta t$ the timestep used per integration step. Different colours correspond to different values for the initial velocity $v_{\rm init}$, as indicated. Results are plotted as a function of the effective value of $\zeta = 0.001 \times 2^n$, which reflects the expectation value for the {\it maximum} number of scatter events per time step. The vertical dotted line corresponds to $\zeta = 0.5$ which is the value adopted in our $N$-body simulations. The bottom panels plot the ratio $\Gamma_\rmn/\Gamma_0$ of the scatter rates averaged over the period $\Delta T$. Errorbars, computed from the variance among ten realisations, are smaller than the symbols shown.}
\label{fig:scatstat}
\end{figure}

Each time step $\Delta t$ we proceed as in our $N$-body code: we draw a number of scattering events $N_{\rm scat}$ from a Poisson distribution with a mean of $\langle N_{\rm scat} \rangle$, and scatter particle $P$ a total of $N_{\rm scat}$ times against particles drawn at random from the MB distribution, each time updating the velocity of $P$. Each subsequent time step, we recompute $\langle N_{\rm scat} \rangle$ using the updated value for $v_\rmP$, which enters in the calculation of $\langle v_{\rm rel} \rangle$. At the end of the $N_{\rm step}$ time steps, we register the final velocity of $P$ as well as the total number of scatter events it experienced. Note that $\Delta t$ is based on the value of $\langle N_{\rm scat} \rangle$ at the initial time step, when $v_\rmP = v_{\rm init}$. It is then kept fixed during the entire period $\Delta T$.

For each value of $n$, we repeat this experiment a total of $5 \times 10^6$ times, each time starting $P$ with the same $v_{\rm init}$. The blue histogram in the top-left panel of Fig.~\ref{fig:multiscat} shows the resulting distribution of $v_\rmP$ for $n=0$ and $v_{\rm init} = 3 \sigma$. The grey-shaded histogram shows the MB distribution of the heat bath and is shown for comparison. Note the large spike at $v_\rmP = v_{\rm init} = 3\sigma$, which represents all realisations in which $P$ did not scatter throughout the entire period $\Delta T$. The blue histogram in the bottom-left panel shows the distribution of $N_{\rm scat}$, the number of scatter events per small time step $\Delta t$, in all $5 \times 10^6$ realisations. Since for $n=0$ the time step $\Delta t$ is tuned so that the initial $\langle N_{\rm scat} \rangle = 0.001$, the vast majority of time steps have $N_{\rm scat} = 0$ or $1$; the fraction of time steps with $N_{\rm scat} \geq 2$ is only $2.4 \times 10^{-6}$. The average number of scatter events experienced by $P$ per time step $\Delta t$ is $0.00088$, which is somewhat smaller than the initial expectation value of $0.001$. The reason for this difference is that as $P$ thermalizes, its average velocity approaches that of the MB distribution, which is smaller than $v_{\rm init}$. Hence, as $P$ thermalizes, $\langle v_{\rm rel} \rangle$ decreases, which reduces its probability of scattering. 

The orange, violet and ocre histograms in Fig.~\ref{fig:multiscat} show the results for $n=6$, $9$ and $11$, corresponding to $N_{\rm step}=64$, $4$ and $1$, respectively. Recall that in each case $P$ is integrated for the same total duration $\Delta T$, but that the time step used, and thus the effective value of $\zeta$, increases with $n$ as indicated in each panel. Note that the distributions of the final velocities (top panels) are virtually indistinguishable from those for $n=0$. Yet, the lower panels show that the corresponding distributions for $N_{\rm scat}$, the number of scatter events per time step $\Delta t$, are very different, with averages of $\langle N_{\rm scat} \rangle = 0.028$, $0.465$, and $2.047$, respectively. However, the average number of scatter events experienced over the entire time interval $\Delta T$ remains fairly similar; it slightly increases from $1.802$ in the case of $n=0$ to $2.047$ in the case where $n=11$. This is a consequence of the fact that when $\Delta t$ is too large, the scattering probabilities are not properly updated based on the evolving velocity of $P$. Hence, advocates of the `standard model' are correct that allowing multi-scattering per particle per time step can cause a systematic error in the scatter frequency. However, unless the values for $\langle N_{\rm scat} \rangle$ per time step are allowed to be excessively large (which we quantify below), the impact of this systematic has a negligible impact. The first hint that this is indeed the case is provided by the top panels of Fig.~\ref{fig:multiscat}, which shows that the PDF for $v_\rmP$ appears invariant to increasing $\langle N_{\rm scat} \rangle$, and thus $\zeta$, by over three orders of magnitude, from $0.001$ to $2.048$.
\begin{figure}
\centering
\includegraphics[width=0.46\textwidth]{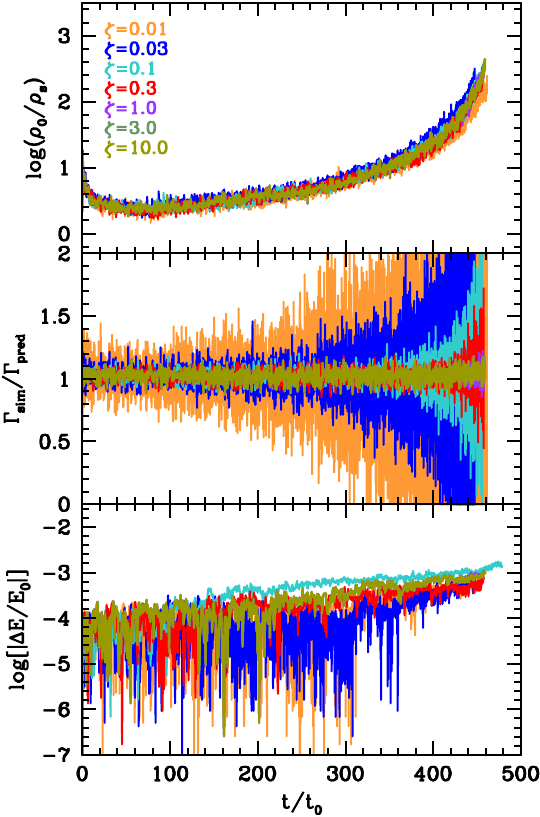}
  \caption{The impact of varying $\zeta$ on $N$-body simulations of the gravothermal evolution of the NFWc20$^{*}$ halo simulated with $N_\rmp=10^5$ particles. Each simulation uses a cross section $\sigma_\rmm = 50\cmg$, starts from the same initial conditions, and uses a Plummer softening of $\epsilon = 0.025$. Top panel shows the evolution of the core density, middle panel shows the ratio of the scatter rate in the simulation to the expected scatter rate computed using equation~(\ref{GammaExp}), and the bottom panel shows the level of energy conservation. Note that changing $\zeta$ by two orders of magnitude has no significant impact on the simulation results, except for the level of noise in $\Gamma_{\rm sim}/\Gamma_{\rm pred}$, which increases with descreasing $\zeta$ due to the fact that smaller $\zeta$ implies smaller time steps. }
\label{fig:kappa}
\end{figure}

To better quantify this, the top panel of Fig.~\ref{fig:scatstat} plots the ratio $\langle v_\rmP \rangle_\rmn / \langle v_\rmP \rangle_0$, where the subscript indicates the value of $n$ used in the experiment, as a function of the effective value of $\zeta = 0.001 \times 2^n$. The vertical dotted line corresponds to $\zeta = 0.5$, which is the value we have adopted in our simulations. Different colours correspond to different values for the initial velocity $v_{\rm init}$, as indicated. As is evident, $\langle v_\rmP \rangle$ is extremely robust to changes in $\zeta$, and we have verified that this also holds for the 5, 50 and 95 percentiles of the full $v_\rmP$ distributions. The bottom panels plot the ratio $\Gamma_\rmn/\Gamma_0$ of the scattering rates, averaged over the period $\Delta T$. Note that $\Gamma_\rmn$ converges to a constant value when $\zeta$ is sufficiently small. However, when $\zeta$ is too large it causes an over(under)-estimate of the scattering rate for particles that are moving faster (slower) than the average. Note, though, that the results shown here are for particle beams of a single initial velocity. When averaged over an entire distribution of particle velocities, the error in the {\it total} scattering rate will be significantly smaller. In addition, in the $N$-body simulations $\zeta$ reflects the expectation value for the {\it maximum} number of scattering events per time step, whereas in the idealised experiment above we have assumed that all particles have the same $\langle N_{\rm scat} \rangle$. 

Finally, to explicitly demonstrate that large values of $\zeta$ are not detrimental to the simulation outcome, we have run the NFWc20$^{*}$ halo using different values of $\zeta$. Fig.~\ref{fig:kappa} shows the evolution of the central density of our NFWc20$^{*}$ halo with a collisional cross section of $\tilde{\sigma}_\rmm = 0.375$, run with various values of $\zeta$, as indicated. Note that here we have used simulations with $N = 10^5$ particles only, as runs with small $\zeta$ take extremely long to complete (the $\zeta = 0.1$ run took $244$ hours on a single processor, compared to $34$ hours for our fiducial value of $\zeta=0.5$). The top panel shows the evolution of the central density, which shows almost no dependence on $\zeta$. The middle panel shows the ratio of the scattering rate in the simulation, $\Gamma_{\rm sim}$, to the predicted scattering rate based on the instantaneous density and velocity dispersion profile (see Section~\ref{sec:compA}). Within the noise, this ratio is consistent with unity at all times and for all values of $\zeta$ probed. Note that $\Gamma_{\rm sim}$ is inferred from the number of scattering events integrated over a period of ten simulation time steps. Since smaller values of $\zeta$ imply smaller time steps, the Poisson noise increases with decreasing $\zeta$. The bottom panel plots the evolution of $\vert \Delta E/E_0 \vert = \vert[E(t) - E(0)]/E(0)\vert$ where $E(t)$ is the total energy of the system at time $t$. Note that energy conservation is typically at the level of $10^{-4}$, again independent of the value of $\zeta$. 

\section{Realisation Noise}
\label{App:realisation}
\begin{figure*}
\centering
\includegraphics[width=0.8\textwidth]{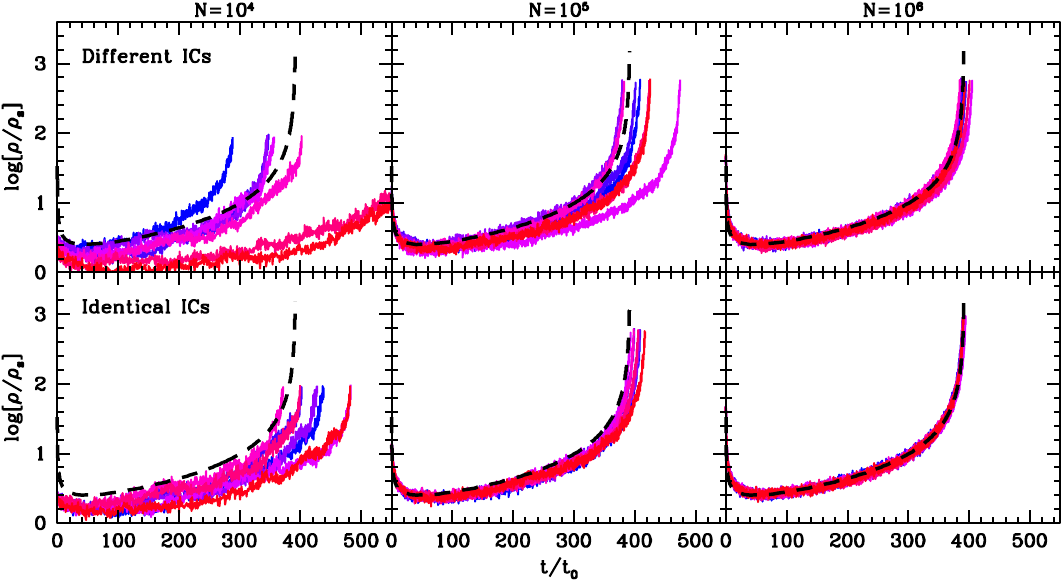}
  \caption{The impact of mass resolution on modelling core collapse with the \SphMC code. From left to right, different panels show results for different $N_\rmp$ ranging from $10^4$ (left) to $10^6$ (right). Each panel shows the results for 7 different simulations of the same NFWc20$^{*}$ halo using $\sigma_\rmm = 25 \cmg$ ($\tilde{\sigma}_\rmm = 0.19$). In the top panels, each simulation starts from a different IC file, while the bottom panels use an identical IC file, but a different seed for the random number generator used to model particle scattering.  Black, dashed curves show the prediction from the \GTF code using $(\alpha,\beta) = (0.50,0.90)$ and are shown for comparison.}
\label{fig:massres}
\end{figure*}

Each simulation of core collapse, be it with the \SphMC code or with an $N$-body code, is subject to two sources of realisation noise. The first arises from the specific initial conditions that are set up by sampling particles from the system's distribution function as described in Section~\ref{sec:ICs}. The second comes from the scattering process, which requires drawing several random numbers to decide which pairs of particles are going to scatter and to pick a random scattering angle for those that do. Here we investigate the impact of both sources of realisation noise. 

We construct a suite of random realisations of ICs for the NFWc20$^{*}$ halo (see Table~1). We consider three different mass resolutions: $N_\rmp=10^4$, $10^5$ and $10^6$, and for each of these we construct 7 different realisations using different random seeds. We run each of these 24 simulations with the \SphMC code, always with the same numerical parameters and with a collision cross section $\sigma_\rmm = 25 \cmg$ ($\tilde{\sigma}_\rmm = 0.19$).

The results are shown in the upper panels of Fig.~\ref{fig:massres}. Different panels correspond to different mass resolutions, while different colours correspond to different IC realisations. The black dashed curves show the  prediction from the \GTF code using $(\alpha,\beta) = (0.50,0.90)$ and are shown for comparison. For $N_\rmp \gta 10^6$ the impact of realisation noise is modest at best, resulting in a $\sim 2$ percent standard deviation in the collapse time. This is small enough for an accurate calibration of the GTF conductivity parameters $\alpha$ and $\beta$. However, for $N_\rmp \leq 10^5$ realisation noise causes a large divergence in collapse times, ranging from core collapse that is too early to cases where core collapse is drastically delayed. These findings are in excellent agreement with \citet{Mace.etal.24}, who found very similar behaviour for $N$-body simulations. 

In the bottom panels, the ICs of all simulations are identical, but different random seeds are used to sample the scattering events. As is evident, this results in a significantly reduced level of stochastisticity, indicating that the realisation noise that arises from the ICs is the dominant source of stochasticity. 

\section{Validation of \SphMC code}
\label{App:validation}
\begin{figure*}
\centering
\includegraphics[width=0.8\textwidth]{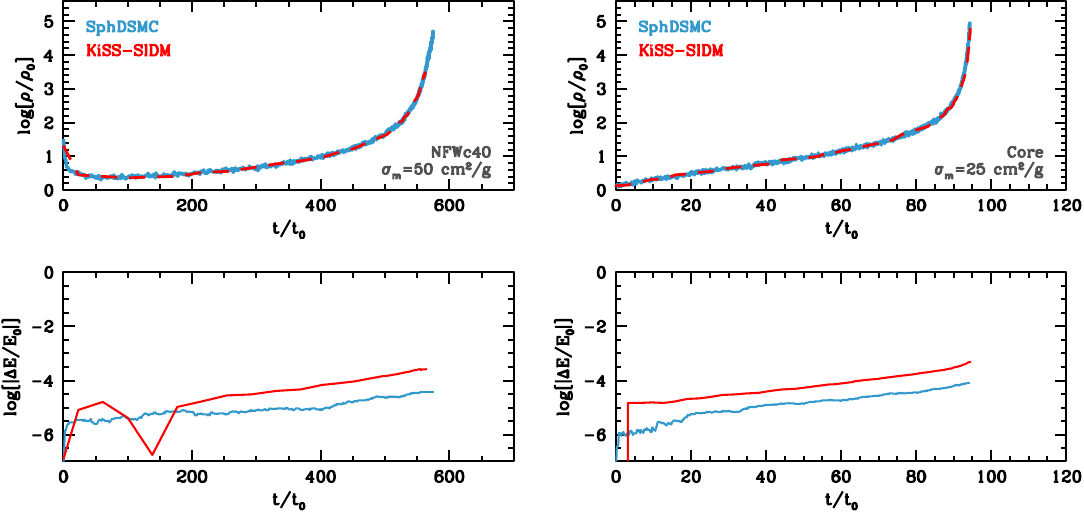}
  \caption{Comparison of the results obtained with \SphMC (azure-blue) and the {\tt KiSS-SIDM} code (red) of \citet{Gurian.May.25} for the NFWc40 halo with $\sigma_\rmm = 50 \cmg$ (left panels) and the \Core halo with $\sigma_\rmm = 25 \cmg$ (right panels). The {\tt KiSS-SIDM} simulations were kindly run on our ICs by J. Gurian. Top panels show the evolution of the central density, while bottom panels show the level of energy conservation. Note that the results are in excellent agreement.}
\label{fig:compGUR}
\end{figure*}

In order to validate the \SphMC code described in Section~\ref{sec:SphMC} we compare its results to those obtained with the {\tt KiSS-SIDM} code described in \citet{Gurian.May.25}. {\tt KiSS-SIDM} is conceptually very similar to \SphMC in that it also uses the DSMC method of \citet{Bird.94} for the scattering, but there are small differences. For example, {\tt KiSS-SIDM} does not use LC regularization for the orbit integration and does not use a remainder-accumulator for the collisions. Fig.~\ref{fig:compGUR} compares the results obtained with the \SphMC code (in azure-blue) for two different halos and cross sections (as indicated) with those obtained with {\tt KiSS-SIDM} (in red). The latter are provided by J. Gurian who kindly ran his code over our ICs. As is apparent, the agreement is excellent.

\bigskip
\bigskip



\label{lastpage}

\end{document}